\documentclass[a4paper,11pt]{article} 

\usepackage{graphicx,amssymb,color,amsthm,amsmath}
\usepackage{longtable}

\usepackage{authblk} 
\usepackage{tikz}
\usepackage{ulem}
\usetikzlibrary{shapes.geometric, arrows}
\usetikzlibrary{calc}
\usepackage{epsfig,float}
\usepackage{color}
\usepackage{subfigure}
\usepackage{float}
\usepackage{graphicx}
\usepackage{pdflscape}
\usepackage{caption}
\usepackage{verbatim}

\usepackage[colorlinks=true,citecolor=blue, linkcolor=blue, urlcolor=blue]{hyperref}
\allowdisplaybreaks
\tikzstyle{start1} = [rectangle, rounded corners, minimum width=3cm, minimum height=1cm, text centered, draw=black, fill=red!30]
\tikzstyle{decision1} = [diamond, minimum width=3cm, minimum height=1cm, text centered, draw=black, fill=orange!30]
\tikzstyle{yes1} = [rectangle, minimum width=3cm, minimum height=1cm, text centered, draw=black, fill=green!30]
\tikzstyle{no1} = [rectangle, minimum width=3cm, minimum height=1cm, text centered, draw=black, fill=blue!30]
\tikzstyle{arrow1} = [thick,->,>=stealth]
\tikzstyle{start2} = [rectangle, rounded corners, minimum width=3cm, minimum height=1cm, text centered, draw=black, fill=red!30]
\tikzstyle{decision2} = [diamond, minimum width=3cm, minimum height=1cm, text centered, draw=black, fill=orange!30]
\tikzstyle{yes2} = [rectangle, minimum width=3cm, minimum height=1cm, text centered, draw=black, fill=green!30]
\tikzstyle{no2} = [rectangle, minimum width=3cm, minimum height=1cm, text centered, draw=black, fill=blue!30]
\tikzstyle{factor2} = [ellipse, minimum width=2.5cm, minimum height=1cm, text centered, draw=black, fill=yellow!30]
\tikzstyle{arrow2} = [thick,->,>=stealth]
\tikzstyle{dottedarrow2} = [thick,->,>=stealth, dotted]

\allowdisplaybreaks
\title{ Mathematical Modeling, Analysis and Simulation Utilizing Machine Learning Tools for Assessing the Impact of Climate Lobbying }

\begin{document}

\author[1]{Andrew Jacoby}
\author[2]{Samiran Ghosh}

\author[3]{Malay Banerjee}
\author[4,*]{Aditi Ghosh}
\author[5]{Padmanabhan Seshaiyer}

\affil[1]{ Massachusetts Institute of Technology, Department of Mathematics}
\affil[2]{Indian Institute of Technology Bombay, Department of Mathematics}
\affil[3]{Indian Institute of Technology Kanpur, Department of Mathematics and Statistics}
\affil[4, *]{East Texas A\&M University, Department of Mathematics}
\affil[5]{George Mason University, Department of Mathematics}

\affil[*]{Aditi.Ghosh@tamuc.edu}

\maketitle
\begin{abstract}

Climate policy and legislation has a significant influence on both domestic and global responses to the pressing environmental challenges of our time. The effectiveness of such climate legislation is closely tied to the complex dynamics among elected officials, a dynamic significantly shaped by the relentless efforts of lobbying. This project aims to develop a novel compartmental model to forecast the trajectory of climate legislation within the United States. By understanding the  dynamics surrounding floor votes, the ramifications of lobbying, and the flow of campaign donations within the chambers of the U.S. Congress, we aim to validate our model through a comprehensive case study of the American Clean Energy and Security Act (ACESA). Our model adeptly captures the nonlinear dynamics among diverse legislative factions, including centrists, ardent supporters, and vocal opponents of the bill, culminating in a rich dynamics of final voting outcomes. We conduct a stability analysis of the model, estimating parameters from public lobbying records and a robust body of existing literature. The numerical verification against the pivotal 2009 ACESA vote, alongside contemporary research, underscores the model’s promising potential as a tool to understand the dynamics of climate lobbying. We also analyse the pathways  of the model that aims to guide future legislative endeavors in the pursuit of effective climate action.

\end{abstract}

\section{Introduction}

Policy serves as a crucial tool for both domestic and global climate action. The efficacy of measures aimed at combating the dire challenge of climate change is closely related to the collaborative efforts of elected officials trying to enact a comprehensive environmental legislation. Unfortunately, the United States has historically fallen short in enacting significant climate policies. Current climate goals are unattainable without substantial legislative action. The Paris Agreement, with its noble intent, introduced a framework of voluntary emission reduction commitments, aspiring to cap global warming below the critical thresholds of 1.5 degrees Celsius and 2.0 degrees Celsius.  However, the agreement failed to yield significant results, and the urgent action needed to maintain the 1.5-degree goal did not occur \cite{stoddard2021}. Despite this, numerous opportunities for domestic climate legislation have arisen.

Between 2003 and 2007, several climate bills were proposed, gaining some traction, but none successfully passed into law \cite{c2es2024}. Notably, the Climate Stewardship Act was introduced in the 108th, 109th, and 110th Congresses but failed to secure passage. This bill aimed to establish a cap-and-trade system for carbon emissions \cite{c2es2024}. Progress was made in 2007 when Congress, through the Consolidated Appropriations Act, mandated the EPA to require public disclosure of emissions from major sources. This led to the creation of the Greenhouse Gas Reporting Program database, which provided nationwide emissions data \cite{c2es2024}.

From 2008 to 2010, additional cap-and-trade proposals were introduced, with the American Clean Energy and Security Act (ACESA) passing the House but ultimately failing in the Senate due to the filibuster threat \cite{meng2019}. Legislative momentum around climate change waned between 2010 and 2018, though some efforts continued, including renewable energy tax credits, funding for carbon capture research, and the introduction of carbon pricing bills. Additionally, the bipartisan Climate Solutions Caucus was established during this period \cite{c2es2024}.

In 2019, Congress renewed its focus on climate action with the introduction of the Green New Deal, the formation of a Select Committee on the Climate Crisis in the House, a bipartisan climate solutions caucus in the Senate, and market-based mitigation efforts \cite{c2es2024}. By December 2020, Congress passed a legislative package funding clean energy research and development, providing green tax incentives, and mandating the EPA to phase down hydrofluorocarbons over the next 15 years \cite{c2es2024}. While this package laid the groundwork for future legislative efforts under the Biden administration, it fell short of enacting the immediate, impactful actions necessary to significantly reduce greenhouse gas emissions in the United States.

To further investigate the potential impact of earlier climate bills, this article will analyze ACESA, one of the most significant pieces of climate legislation in U.S. history. ACESA introduced a market-based approach to reducing carbon emissions through a cap-and-trade system. Under this system, a limited number of emissions permits would be allocated, allowing major carbon emitters to produce CO\textsubscript{2} up to the amount of their allocated permits. If their emissions were lower than their permits, they could sell the excess permits to other companies \cite{meng2019}. This mechanism effectively set a nationwide cap on CO\textsubscript{2} emissions, which would decrease over time.

This declining cap led to intense lobbying efforts. In fact, lobbying surrounding ACESA accounted for 14\% of all lobbying expenditures during that period \cite{meng2019}. Between the first quarter of 2007 and the first quarter of 2009, the number of lobbyists advocating for and against cap-and-trade legislation increased significantly. Lobbyists representing corporations, the fossil fuel industry, conservative think tanks, and foundations were more successful than pro-environment groups. Their efforts reduced the likelihood of ACESA's passage by approximately 13\% \cite{meng2019}. Combined with the legacy of the Bush administration’s climate resistance within the Republican Party, ACESA ultimately failed to pass the Senate despite clearing the House due to the threat of a conservative filibuster \cite{meng2019}.

At its core, lobbying can be viewed as an interaction between interest groups (lobbyists) and policymakers, where lobbyists attempt to influence decision-making through campaign contributions, information dissemination, and other methods. The authors believe that that mathematical models can be used to represent such lobbying dynamics, especially in the context of political influence and legislation for ACES. Specifically, mathematical modeling can play a significant role in lobbying efforts by providing data-driven insights, predictions, and evidence that can influence policy decisions \cite{aragon2024mathematics}. There have been several mathematical models for lobbying including game-theoretic models that can help estimation of risks along with allocation of costs \cite{ward2004pressure, aragon2024mathematical} and optimization techniques to help lobbyists design policies that maximize desired outcomes while minimizing negative side effects \cite{nehama2015complexity}. Such mathematical, statistical, and computational methods have long been used to predict and analyze roll-call votes in Congress. These approaches utilize a wide range of data, including debate transcripts, bill text, legislator ideology, party composition, lobbying activity, and campaign finance information \cite{Budhwar2018} \cite{Smith2012} \cite{Goldblatt2012} \cite{Karimi2020} \cite{Lan2013} \cite{HenighanKravitz} \cite{KimKunisky2021}. Machine learning (ML) and artificial intelligence (AI) methods, in particular, have been applied to predict voting outcomes, using models such as neural networks and logistic regression \cite{Smith2012} \cite{Bari2021} \cite{Budhwar2018} \cite{Goldblatt2012}. Additionally, network analysis and spatial models capture the complex relationships and ideological distances among legislators \cite{Lan2013} \cite{Poole1985}. Game theory and statistics-based models are also employed to assess the influence of lobbying and other variables on the likelihood of bill passage \cite{meng2019}.

For example, an ensemble approach combining logistic regression, SVM, and neural networks achieved an 80.13\% accuracy rate in predicting bill passage in Congress by utilizing features such as the sponsor’s identity, bill text, and timing \cite{Budhwar2018}. Another study employed a Transformer-based text embedding model alongside campaign finance data, achieving over 90\% accuracy in roll-call vote prediction, underscoring the critical role of financial interests in shaping legislative behavior \cite{Bari2021}. Spatial models have also been instrumental in illuminating legislative behavior \cite{Poole1985}. The NOMINAl Three-step Estimation model (NOMINATE), for instance, places legislators and roll-call votes in a common ideological space, enabling the estimation of ideological positions based on voting patterns \cite{Poole1985}. This model has demonstrated that many roll-call votes can be explained by a single liberal-conservative dimension, although multidimensional analysis can uncover more intricate voting behaviors \cite{Poole1985}.

The authors in this paper believe that mathematical  models that incorporate human behavior can offer a novel and insightful way to study the dynamics of lobbying. By viewing lobbying as a process of influence spreading through a network, we can apply concepts from epidemiology to predict and understand how lobbying efforts can succeed or fail. This approach allows lobbyists, policymakers, and researchers to make more informed decisions about the strategies, timing, and resources needed for effective advocacy campaigns. Combined with Machine Learning (ML) these mathematical models can be highly effective in predicting legislative outcomes with substantial accuracy \cite{raissi2019parameter}.

This article proposes a novel method for understanding the dynamics that influence policy passage. While existing research utilizes spatial modeling, ML, and statistical methods, we propose a novel differential equations-based approach to simulating and predicting roll-call vote outcomes.  \textbf{Our objectives are threefold: (a) to construct a novel and innovative dynamic model for understanding factors that affect roll-call votes in Congress, (b) to investigate how the interactions between these factors shape legislative behavior, and (c) to predict the likelihood of bill passage. } We use a compartmental system of differential equations to model the intricate relationships between lobbyists, legislators, and financial contributions, allowing us to analyze the nonlinear dynamics that determine the fate of bills in Congress and the evolving opinions of legislators over time. The model is validated by applying it to the ACESA, where it successfully predicts roll-call vote outcomes within 5\% of the actual results.

This article introduces a novel compartmental system of differential equations to model legislative outcomes, with a comprehensive stability and sensitivity analysis that improves our understanding of parameter impacts. Using Support Vector Machines (SVM), we identify critical boundaries for bill passage, offering a robust approach to predict legislative success. Through rigorous model construction, positivity and boundedness proofs, and targeted simulations applied to the American Clean Energy and Securities Act, this study provides valuable insights into the dynamics of policy enactment, setting a foundation for future research in legislative decision-making modeling. IN section 2, we discuss the materials and methods followed by Stability Analysis and section 3. Sensitivity Analysis and Results are discussed in Section 4 and 5 respectively. Section 6 concludes with Discussion.

\section{Materials and Methods}

Let us consider a legislative body composed of \( S \) voting members. As these individuals are introduced to a bill, each forms an initial opinion. Based on their pre-existing beliefs and ideologies, some legislators will immediately decide to vote in favor of the bill, while others will decide to vote against it. Those in favor of the bill will move into the \( Y \) compartment, and those opposed will move into the \( N \) compartment. However, some legislators will remain undecided, forming the \( C \) compartment. This is visualized using a flow diagram in Fig. \ref{basicflowchart}. The details of the parameters are given in the Appendix in Table \ref{Tab1}.

\begin{figure}[ht!]
    \centering
    \includegraphics[width=10.5cm]{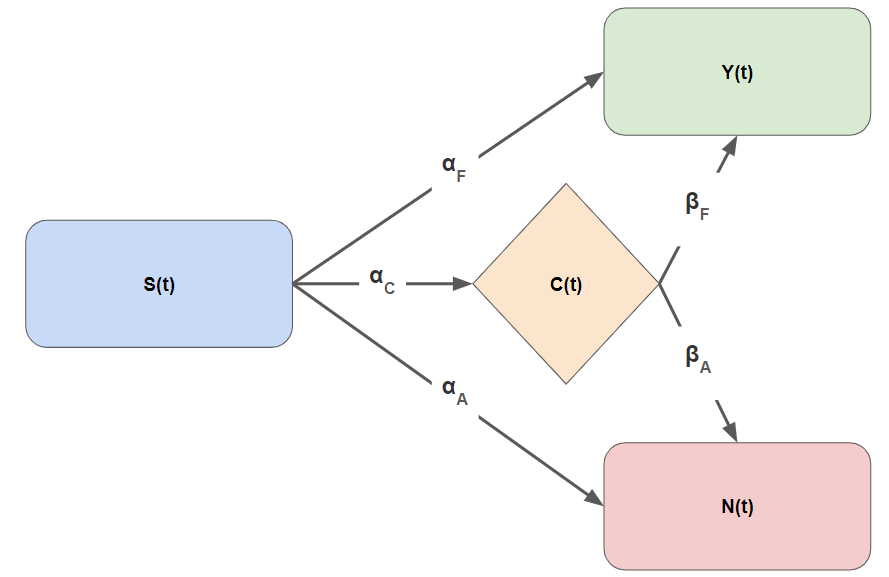}
    \caption{Schematic diagram of the model without including the influence of lobbying and donations}
    \label{basicflowchart}
\end{figure}

In the {baseline} model, undecided legislators are influenced by peer pressure and the opinions of their colleagues, leading them to eventually join either the \( Y \) or \( N \) compartments. This framework can be represented by a system of differential equations:

\begin{subequations}\label{base_system}
\begin{eqnarray}
S'(t)&=&-(\alpha_F+\alpha_A+\alpha_C)S(t), \\
C'(t)&=& \alpha_C S(t)-\beta_F C(t)Y(t) -\beta_A C(t)N(t), \\
Y'(t)&=& \alpha_F S(t)+\beta_F C(t)Y(t), \\
N'(t)&=& \alpha_A S(t)+\beta_A C(t)N(t).
\end{eqnarray}
\end{subequations}

Here, \( \alpha_F \), \( \alpha_A \), and \( \alpha_C \) represent the proportion of legislators in \( S \) who are initially inclined to support, oppose, or remain undecided about the bill, respectively. The parameters \( \beta_F \) and \( \beta_A \) denote the rates at which undecided legislators (\( C \)) are persuaded to vote in favor (\( Y \)) or against (\( N \)) the bill due to interactions with their peers.

While peer influence is a critical factor, it is not the only determinant in the decision-making process of undecided legislators. Lobbying and potential campaign contributions also exert significant influence.

\subsection{Lobbying and Firms}

To evaluate the influence of lobbying, we examine the potential financial impact that the bill would have on firms. Following the methodology in \cite{meng2019}, let \( \pi_i^0 \) denote the value of firm \( i \) in the absence of the bill, and \( \pi_i^* \) its value under the bill where i ranges from 1 to N where N is the total number of firms. The change in firm value resulting from the bill is expressed as:
\[
\Delta \pi_i = \pi_i^* - \pi_i^0.
\]
This change, \( \Delta \pi_i \), is a function of various parameters related to the bill. If \( \Delta \pi_i > 0 \), the firm will lobby in favor of the bill and fall into the $N_F$ category; if \( \Delta \pi_i < 0 \), the firm will lobby against it and fall into the $N_A$ category. As a result, all firms have either a positive or negative impact on the bill.

Suppose the maximum total number of lobbyists is \( L \), which is constant over time. These lobbyists will divide into two groups: \( L_F \) (lobbying in favor) and \( L_A \) (lobbying against). At a given time \( t \), the total amount of money available for lobbying in favor of the bill is:
\[
M^{total}_{F} = \sum_{i=1}^{N_F} \rho_Y \Delta \pi_i,
\]
and the total amount of money available for lobbying against the bill is:
\[
M^{total}_{A} = \sum_{i=1}^{N_A} \rho_N \Delta \pi_i.
\]

Here, \( \rho_Y \) represents the proportion of potential financial gains that a firm allocates to lobbying for the bill, while \( \rho_N \) represents the proportion of potential financial losses allocated to lobbying against it.

The amount of money allocated to lobbyists in favor of the bill is:
\[
M_{LF} = (1-\psi) M^{total}_{F},
\]
and the amount allocated to lobbyists against the bill is:
\[
M_{LA} = (1-\psi) M^{total}_{A}.
\]

At time \( t \), the number of lobbyists supporting or opposing the bill is \( L_F(t) \) and \( L_A(t) \), respectively. The number of lobbyists who have not yet taken a position is \( L - L_F(t) - L_A(t) \). These lobbyists will join either \( L_F \) or \( L_A \) compartments at rates proportional to \( M_F \) or \( M_A \), respectively. The differential equations governing the number of lobbyists are:
\[
\dot{L_F} = \nu_F M_{LF} (L - L_F(t) - L_A(t)),
\]
\[
\dot{L_A} = \nu_A M_{LA}(L - L_F(t) - L_A(t)),
\]
where \( L \geq L_F(0), L_A(0) \geq 0 \), and \( \nu_F \) and \( \nu_A \) represent the rates at which lobbyists are either hired to or decide to work on the ACESA bill.

The amounts of money directly spent on campaigns and neutral legislators in favor of the bill (\( M_F \)) and against it (\( M_A \)) evolve according to:
\[
\dot{M_F} = -I_F(t),
\]
\[
\dot{M_A} = -I_A(t).
\]

When \( M_F = 0 \), \( I_F(t) = 0 \), and similarly, when \( M_A = 0 \), it implies \( I_A(t) = 0 \). The terms \( I_F(t) \) and \( I_A(t) \) represent the money spent on donations directly to candidates, Political Action Committees (PACs), and Super-PACs.

\subsection{{Enhanced}  Model}

By incorporating lobbying and campaign contributions, we extend the {baseline} model to obtain {the flow diagram for the extended model illustrated in Fig.} \ref{fullflowchart}:

\begin{figure}[ht!]
    \centering
    \includegraphics[width=10.5cm]{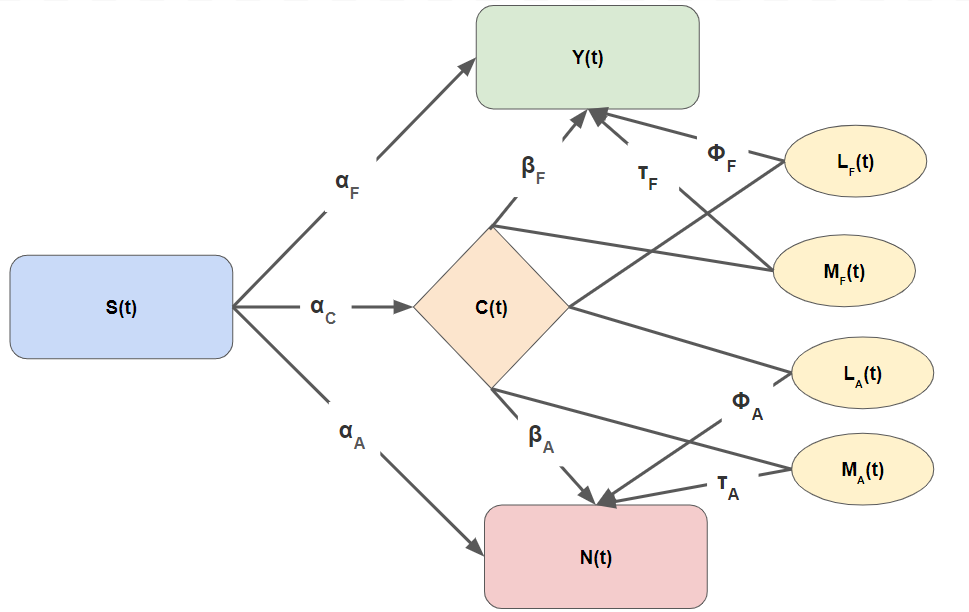}
    \caption{Schematic diagram of the model including the impact of lobbying and donations}
    \label{fullflowchart}
\end{figure}


The {extended model may be described} by the following set of differential equations, for details of the parameters  and its values, please refer to the Appendix in Table \ref{Tab1}.

\begin{subequations}\label{main_system_1}
\begin{eqnarray}
S'(t) &=& -(\alpha_F + \alpha_A + \alpha_C) S(t), \\
L_F'(t) &=& \nu_F M_{LF}(L - L_F(t) - L_A(t)), \\
L_A'(t) &=& \nu_A M_{LA}(L - L_F(t) - L_A(t)), \\
M_F'(t) &=& -I_F(t), \\
M_A'(t) &=& -I_A(t), \\
C'(t) &=& \alpha_C S(t) - \beta_F C(t)Y(t) - \beta_A C(t)N(t) - \phi_F L_F(t) C(t) - \phi_A L_A(t) C(t) \nonumber\\
&&- (\tau_F I_F(t) + \tau_A I_A(t))C(t), \\
Y'(t) &=& \alpha_F S(t) + \beta_F C(t)Y(t) + \phi_F L_F(t) C(t) + \tau_F I_F(t) C(t), \\
N'(t) &=& \alpha_A S(t) + \beta_A C(t)N(t) + \phi_A L_A(t) C(t) + \tau_A I_A(t) C(t).
\end{eqnarray}
\end{subequations}

\subsection{Positivity and Boundedness}

In order to show the well-posedness of the model it is critical that we show that the model is both positive and bounded. Positivity ensures that the solutions remain within realistic, interpretable bounds, while boundedness prevents erratic behavior and indicates long-term stability. These properties enhance the reliability and applicability of the model in its intended context, providing confidence in analysis and predictions derived from it.

\subsubsection{Positivity}
We begin by proving the positivity of the lobbying equations separately. 
\begin{eqnarray}
    \frac{dL_F}{dt} &=&\nu_F M_{LF} (L- L_F - L_A),\\
    \frac{dL_A}{dt} &=&\nu_A M_{LA} (L- L_F - L_A).
\end{eqnarray}
These equations have the conditions that $L> L_F(0) \geq 0$ and $L> L_A(0) \geq 0$. Also, note that $\nu_F M_{LF}>0$ and $\nu_A M_{LA}>0$, and, if for some time $t_*>0$, $\frac{dL_F}{dt}|_{t_*}=0$ implies $\frac{dL_A}{dt}|_{t_*}=0$. Consequently, $\frac{d^nL_F}{dt^n}|_{t_*}=0$ implies $\frac{d^nL_A}{dt^n}|_{t_*}=0$, for all $n \in \mathbb{N}$.

Now if possible, consider a situation where $L_F(t)<0$ for some $t \in [0,\infty)$. In order for this to occur, there must exist some time $a \in [0,t)$ such that $L_F(a) = 0, L_F'(a)<0$. Without loss of generality, assume the chosen $a$ is the first time satisfying the previous conditions. Consequently, $\frac{dL_F}{dt}|_{t=a} =\nu_F M_{LF} (L - L_A(a))<0,$ implies, $ L_A(a) > L$. That means that  $\exists$ some time $b \in [0,a)$ such that $L_A(b) = L$ and $L_A'(b)>0$. However, that would mean $\frac{dL_A}{dt}|_{t=b} =\nu_A M_{LA} (L- L_F(b) - L)>0$ so $\nu_A M_{LA} (- L_F(b) )>0$ and given that $\nu_a, M_{LA} \in \mathbb{R}_{0^+}$ this creates a contradiction because $L_F(b)<0$, where, $b <a$. Therefore, $L_F(t)\geq 0, t\in [0,\infty)$. Similar methods can be used for the positivity of $L_A$.\\\\
To prove the positivity of the rest of \ref{main_system_1}, we use the following Lemma:\\
\textbf{Lemma:} Suppose $\Omega \subset \mathbb{R} \times \mathbb{R}^n$ is open and $f_i \in C(\Omega, \mathbb{R})$ for $i = 1, 2, \ldots, n$. If for all $x_i = 0$ and $X_t \in \mathbb{R}^n_{0^+}$, $[ f_i|_{x_i=0, X_t \in \mathbb{R}^n_{0^+}} \geq 0, ]$
then $\mathbb{R}^{n_0^+}$ is the invariant domain of the system of equations
$[ \dot{x}_i = f_i(t, x_1, \ldots, x_n), \quad t \geq \tau, \quad i = 1, \ldots, n, ]$
where $(\mathbb{R}^n_{0^+})$ represents the non-negative orthant of $(\mathbb{R}^n)$.\\ \\
Now, from the system of equations \ref{main_system_1} we set, $X_t = (S(t), M_F(t), M_A(t),  C(t), Y(t), N(t)) \in \mathbb{R}_{0+}^6$ and 
\begin{subequations}
 \begin{eqnarray} 
\left.\frac{dS}{dt}\right|_{S = 0, X(t) \in \mathbb{R}_{0^+}^6}&=& 0, \\
\left.\frac{dM_F}{dt}\right|_{M_F = 0, X(t) \in \mathbb{R}_{0^+}^6}&=& 0,\\
\left.\frac{dM_A}{dt}\right|_{M_A = 0, X(t) \in \mathbb{R}_{0^+}^6}&=& 0,\\
\left.\frac{dC}{dt}\right|_{C = 0, X(t) \in \mathbb{R}_{0^+}^6}&=& \alpha_C S \geq 0,\\
\left.\frac{dY}{dt}\right||_{Y = 0, X(t) \in \mathbb{R}_{0^+}^6} &=& \alpha_F S +\phi L_F C+\tau_F I_F(t) C \geq 0, \\
\left.\frac{dN}{dt}\right||_{N = 0, X(t) \in \mathbb{R}_{0^+}^6} &=& \alpha_A S+ \phi L_A C+\tau_A I_A(t) C \geq 0.
\end{eqnarray}
\end{subequations}
Hence by the previous lemma, $\mathbb{R}_{0^+}^6$ is invariant under the system. This complete the proof of positivity of the complete system \eqref{main_system_1}.

 \subsubsection{Boundedness}

    To prove boundedness for this system of differential equations we break it up into three parts. Both the $M_F$ and $M_A$ functions are bounded by definition. Now, we prove the boundedness of the system
    \begin{eqnarray*}
    \frac{dL_F}{dt} &=&\nu_F M_{LF} (L- L_F - L_A),\\
    \frac{dL_A}{dt} &=&\nu_A M_{LA} (L- L_F - L_A).
\end{eqnarray*}
Define $M_{min}=min\{\nu_F M_{LF},\; \nu_A M_{LA}\},$ $M_{max}=max\{\nu_F M_{LF},\; \nu_A M_{LA}\}$ and $\mathcal{L}(t)=L_F(t)+L_A(t)$. Then from the above two equations we can write
$$\frac{d \mathcal{L}(t)}{dt} \leq M_{max} L -M_{min}\mathcal{L}(t).$$
The above inequality implies
$$\mathcal{L}(t) \leq \mathcal{L}(0)e^{-M_{min}t} + \frac{M_{max}L}{M_{min}} \left( 1- e^{-M_{min}t} \right) \leq \mathcal{L}(0) + \frac{M_{max}L}{M_{min}}.$$
This inequality proves the boundedness of $L_F(t)$ and $L_A(t)$.

Next, let $M(t)=S(t)+C(t)+Y(t)+N(t)$, so $\frac{dM}{dt}=\frac{dS}{dt}+\frac{dC}{dt}+\frac{dY}{dt}+\frac{dN}{dt}=0$ then by integrating both sides, we find that the $M(t)=constant$ and therefore the complete system is bounded.

\subsection{Data Source}
\begin{figure}[ht!]
    \centering
\includegraphics[width=10.5cm]{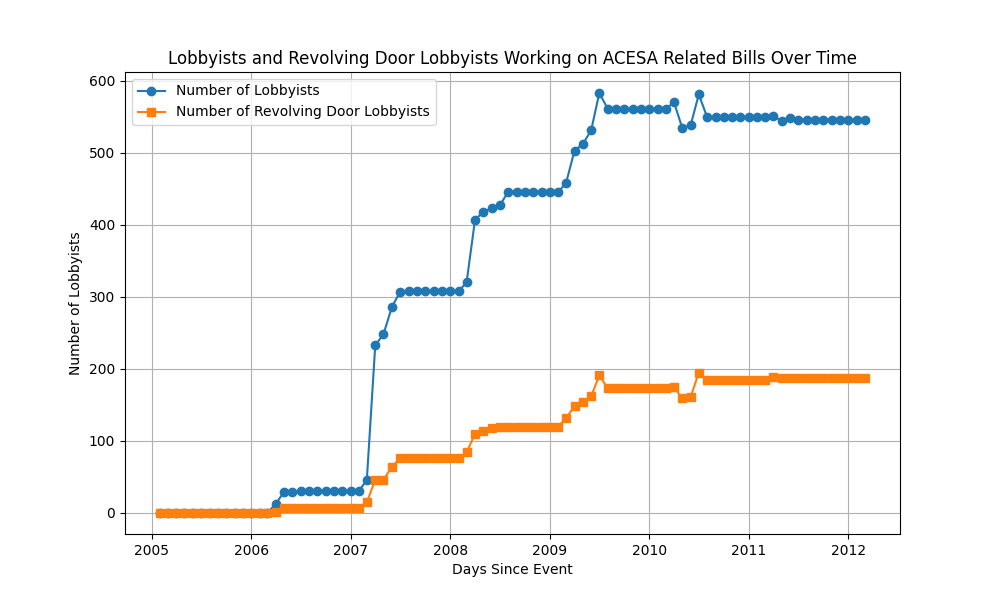} %
\caption{This figure depicts how the number of lobbyists working on ACESA-related legislation changed over time from 2006 to 2012. It includes both regular lobbyists and a subgroup of revolving-door lobbyists.}
\label{Figure1}
\end{figure}

The data utilized for model validation is sourced from ProPublica's lobbying database\cite{Willis2024}, focusing specifically on the American Clean Energy and Security Act (ACESA). This database provides detailed information on the number of lobbyists engaged with ACESA-related issues, including the subset of those who were formerly federally employed\cite{Willis2024}. By examining the timeline and affiliations of these lobbyists, we were able to track the number of individuals with federal employment backgrounds who were involved in lobbying efforts over time. This analysis shed light on the influence that former federal employees might have had on the legislative process, offering a deeper understanding of how lobbying dynamics evolved as the ACESA moved through Congress. The resulting chart tracks the number of lobbyists over time, revealing a pattern similar to logistic growth, with the number of lobbyists increasing from a lower bound of zero to an upper bound of approximately 581 (see Fig. \ref{Figure1}).

\section{Stability Analysis}
To fully understand the behavior of the model and the predictions it makes for the legislative body, we  analyze the stability of  critical points of the model. We aim to determine the end behavior of the model. 

\subsection{Stability of the Lobbying and Donations-Free Model}
We begin by analyzing the stability of the basic model to understand how the system functions without considering the impact of lobbying and donations. 
Consider the corresponding model system:

\begin{subequations}\label{reduced_model}
\begin{eqnarray}
\frac{dS}{dt} &=& -(\alpha_F + \alpha_A + \alpha_C)S, \\
\frac{dC}{dt} &=& \alpha_C S - \beta_F CY - \beta_A CN, \\
\frac{dY}{dt} &=& \alpha_F S + \beta_F CY, \\
\frac{dN}{dt} &=& \alpha_A S + \beta_A CN.
\end{eqnarray}
\end{subequations}

To determine the stability of the equilibrium point \((S,C,Y,N) = (0, C_0, 0, 0)\), we use the transformations \(S(t) = s(t)\), \(C(t) = C_0 + c(t)\), \(Y(t) = y(t)\), and \(N(t) = n(t)\). Substituting these transformations into \eqref{reduced_model} and linearizing the system by neglecting higher-order terms of the small quantities \(s(t)\), \(c(t)\), \(y(t)\), and \(n(t)\), we obtain the following solution (see Sec.\ref{Appendix} for details):

\[
\begin{bmatrix}
    s_1(t) \\
    c_1(t) \\
    y_1(t) \\
    n_1(t)
\end{bmatrix} = \begin{bmatrix}
    s_{10} e^{-(\alpha_F + \alpha_A + \alpha_C)t} \\
    c_{10} \\
    y_{10} e^{\beta_F C_0 t} \\
    n_{10} e^{\beta_A C_0 t}
\end{bmatrix}.
\]

From these solutions, we conclude that as \(t \to \infty\), \(s(t) \to 0\), \(y(t) \to +\infty\), etc. Thus, the equilibrium point \((0, C_0, 0, 0)\) is unstable.
This implies that when the centrist group is dominant ($C_0$) and the "yes" ($Y$) and "no" ($N$) voting groups are negligible, the situation is inherently unstable. Even minor influences, such as campaigns or debates, can disrupt this balance, prompting individuals to take a stance, either in favor of ($Y$) or against ($N$) the bill. As a result, the population is unlikely to remain neutral.

A similar calculation can be used to determine the stability of the equilibrium point \((S, C, Y, N) = (0, 0, Y_0, N_0)\). Using the transformations \(S(t) = s(t)\), \(C(t) = c(t)\), \(Y(t) = Y_0 + y(t)\), and \(N(t) = N_0 + n(t)\), we linearize \eqref{reduced_model} to obtain the following solution ( see  Sec.\ref{Appendix} for details).

\[
\begin{bmatrix}
    s_1(t) \\
    c_1(t) \\
    y_1(t) \\
    n_1(t)
\end{bmatrix} = \begin{bmatrix}
    s_{10} e^{-(\alpha_F + \alpha_A + \alpha_C)t} \\
    c_{10} e^{-(\beta_F Y_0 + \beta_A N_0)t} \\
    y_{10} \\
    n_{10}
\end{bmatrix}.
\]

Transforming back to the original coordinates, we conclude that as \(t \to \infty\), \(s(t) \to 0\), \(c(t) \to 0\), etc. Thus, \((0, 0, Y_0, N_0)\) is stable. From this result, we conclude that as \(t \to \infty\), the \(Y(t)\) and \(N(t)\) functions will converge to positive constants, while the \(S(t)\) and \(C(t)\) functions will approach zero. 

This implies that as time passes and the bill is discussed between the different legislators, the members of the legislature who were either not introduced to the bill ($S$) or undecided ($C_0$) decide to either vote "yes" ($Y$) and "no" ($N$). The legislature's peer pressure and ideology makeup determine if $Y>N$ or $N>Y$ as $t \to \infty$. As a result, the voting decisions of the population are likely to converge.

\subsection{Stability of the Model Including Lobbying and Donations}
To better understand the dynamics of the full system, we now include the impacts of lobbying and campaign donations in the model. As a result, the equilibrium points change. These new equilibrium points can be obtained by solving the following system:

\begin{subequations}\label{full_system}
\begin{eqnarray}
\frac{dS}{dt} &=& -(\alpha_F + \alpha_A + \alpha_C)S(t), \\
\frac{dC}{dt} &=& \alpha_C S - \beta_F CY - \beta_A CN - \phi_F L_F C \nonumber\\
&& - \phi_A L_A C - (\tau_F I_F(t) + \tau_A I_A(t))C, \\
\frac{dY}{dt} &=& \alpha_F S + \beta_F CY + \phi_F L_F C + \tau_F I_F(t) C, \\
\frac{dN}{dt} &=& \alpha_A S + \beta_A CN + \phi_A L_A C + \tau_A I_A(t) C, \\
\frac{dL_F}{dt} &=& \nu_F M_{LF}(L - L_F(t) - L_A(t)), \\
\frac{dL_A}{dt} &=& \nu_A M_{LA}(L - L_F(t) - L_A(t)).
\end{eqnarray}
\end{subequations}

It is important to note that in this system, the \(M_F\) and \(M_A\) functions are not included because the only interaction they have is through \(I_F(t)\) and \(I_A(t)\), which are constants until \(M_F\) and \(M_A\) are greater than or equal to zero. The equilibrium points are \(S=0, C=0, Y=Y, N=N\) and $S=0, C=C,  Y= -\frac{I_F\tau_F - \phi_F L_A + \phi_F L}{\beta_F}, N=-\frac{I_A \tau_A +L_A \phi_A}{\beta_A}$. However, the second point contains negative values so it is outside of the domain of the function. To better perform the stability analysis, we find the analytical solutions to the $L_F(t)$ and $L_A(t)$ functions. 


\subsubsection{Solution to the \texorpdfstring{$L_F(t)$ and $L_A(t)$}{LF(t) and LA(t)} Functions}

Given the differential equations:

\begin{align*}
\frac{dL_F}{dt} &= \nu_F M_{LF}(L - L_F(t) - L_A(t)), \\
\frac{dL_A}{dt} &= \nu_A M_{LA}(L - L_F(t) - L_A(t)),
\end{align*}

we rewrite them in a more suitable form below for analysis.




\begin{align}
\label{subxy}
\frac{dL_F}{dt} &= \nu_F M_{LF} X(t), \\
\frac{dL_A}{dt} &= \nu_A M_{LA} X(t).
\end{align}











It can be shown that (for details of the proof see  Sec. \ref{Appendix} Appendix):



\[
X(t) = (L - L_F(0) - L_A(0)) e^{-(\nu_F M_{LF} + \nu_A M_{LA}) t}.
\]



We integrate  \(L_F(t)\) and \(L_A(t)\) and substitute \(X(\tau)\) to  obtain.









\begin{align*}
L_F(t) &= L_F(0) + \frac{\nu_F M_{LF}}{\nu_F M_{LF} + \nu_A M_{LA}} (L - L_F(0) - L_A(0)) (1 - e^{-(\nu_F M_{LF} + \nu_A M_{LA}) t}), \\
L_A(t) &= L_A(0) + \frac{\nu_A M_{LA}}{\nu_F M_{LF} + \nu_A M_{LA}} (L - L_F(0) - L_A(0)) (1 - e^{-(\nu_F M_{LF} + \nu_A M_{LA}) t}).
\end{align*}

\subsubsection{Full Stability Analysis}

To determine the stability of the equilibrium point \((S, C, Y, N) = (0, 0, Y_0, N_0)\), we use the transformations \(S(t) = s(t)\), \(C(t) = c(t)\), \(Y(t) = Y_0 + y(t)\), and \(N(t) = N_0 + n(t)\) and linearize the system \eqref{full_system} by neglecting higher-order terms of the small quantities \(s(t)\), \(c(t)\), \(y(t)\), \(n(t)\) (see \ref{Appendix} for details). The functions \(L_F(t)\) and \(L_A(t)\) are given by the previously derived solutions to their differential equations:

The solution to the system \eqref{full_system} is:

\[
\begin{bmatrix}
    s_1(t) \\
    c_1(t) \\
    y_1(t) \\
    n_1(t)
\end{bmatrix} = \begin{bmatrix}
    s_{10} e^{-(\alpha_F + \alpha_A + \alpha_C)t} \\
    c_{10} e^{\int \left( -(\beta_F Y_0 + \beta_A N_0 + \phi_F L_F(t) + \phi_A L_A(t) + \tau_F I_F(t) + \tau_A I_A(t)) \right) dt} \\
    y_{10} \\
    n_{10}
\end{bmatrix}.
\]

Despite the complexity of the solution, it can be transformed back to our initial coordinate system. From this, we conclude that as \(t \to \infty\), \(s(t) \to 0\), \(c(t) \to 0\), etc. Thus, \((0, 0, Y_0, N_0)\) is stable. From this result, we conclude that as \(t \to \infty\), the \(Y(t)\) and \(N(t)\) functions will converge to positive constants, while the \(S(t)\) and \(C(t)\) functions will approach zero. This implies that as time passes and the bill is discussed between the different legislators, legislators are lobbied, their campaigns are donated to, or the meet with potential donors, the members of the legislature who were either not introduced to the bill ($S$) or undecided ($C_0$) decide to either vote "yes" ($Y$) and "no" ($N$). The legislature's peer pressure, lobbying, donations and ideology makeup determine if $Y>N$ or $N>Y$ as $t \to \infty$. As a result, the voting decisions of the population are likely to converge.

\section{Sensitivity Analysis}
\begin{figure}[ht!]
    \centering
    \includegraphics[width = 15cm]{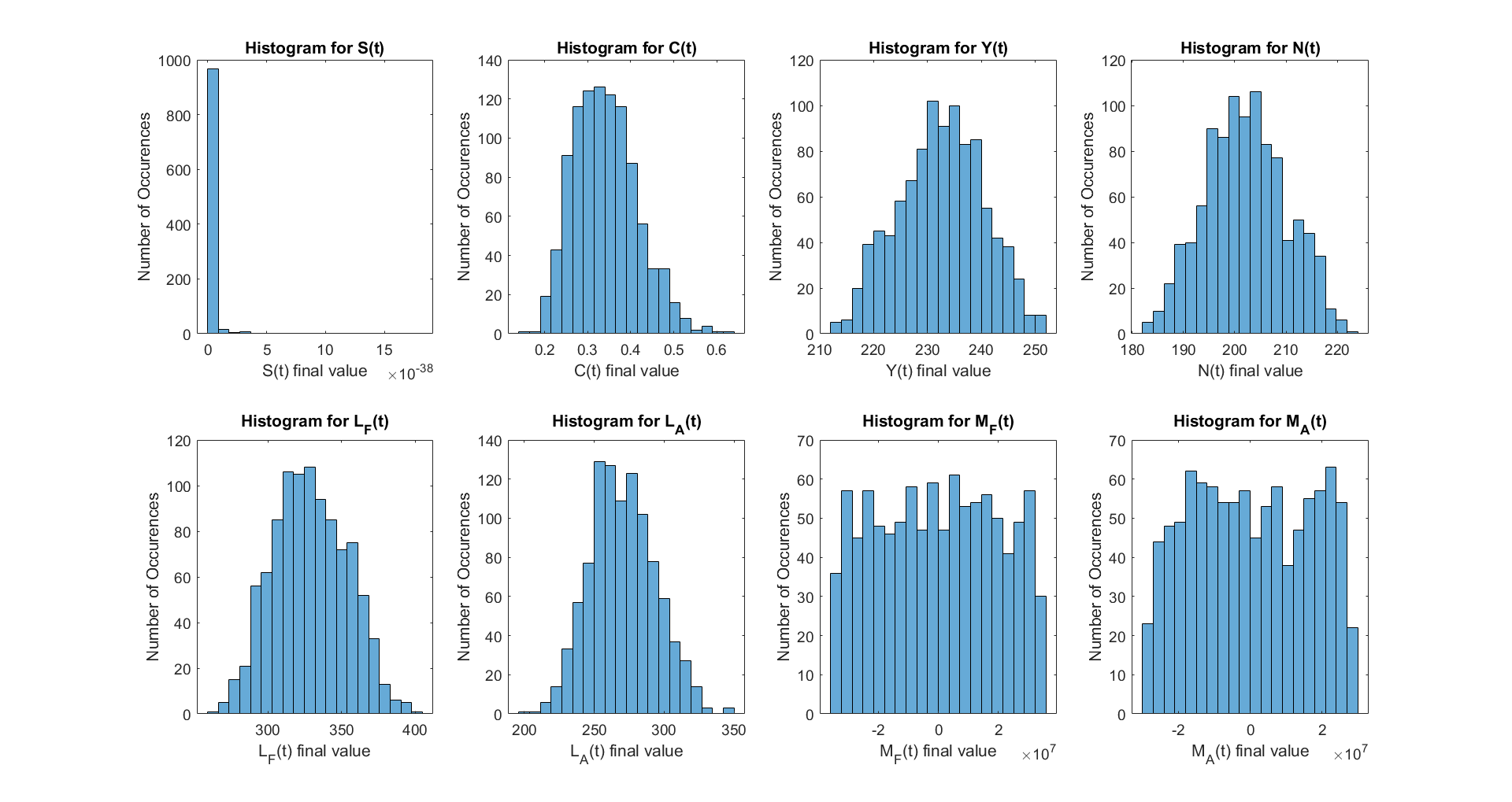}
    \caption{This figure depicts a sensitivity analysis where parameters affecting the flow of centrists into the $Y(t)$ and $N(t)$ compartments are varied by 10\%. The distribution of final values is then recorded and analyzed.}
    \label{Figure8}
\end{figure}

In order to better understand the dynamics of the model, we perform sensitivity analysis by allowing key parameters to vary. Sensitivity analysis is crucial for identifying how changes in model inputs affect the overall outcomes, particularly in systems with multiple interacting variables such as legislative voting dynamics.By varying parameters that control the likelihood of centrists flowing into the $Y(t)$ (Yes) or $N(t)$ (No) compartments by 10\%. Fig.~\ref{Figure8} demonstrates how these changes influence the final distribution of votes. 

For the compartment $S(t)$, representing the initial state of legislators before making a decision, there is very little effect observed when the parameters are varied. This suggests that the total number of legislators participating in the decision process remains stable under these conditions. However, the variables associated with the final opinions of legislators, particularly $C(t)$ (centrists), $Y(t)$ (Yes votes), $N(t)$ (No votes), and the efforts of lobbyists $L_F(t)$ (lobbyists for) and $L_A(t)$ (lobbyists against), show more significant variability. This implies that the model is highly sensitive to changes in certain parameters, especially those controlling centrists’ flow between decision-making states.

\begin{figure}[ht!]
    \centering
    \includegraphics[width = 15cm]{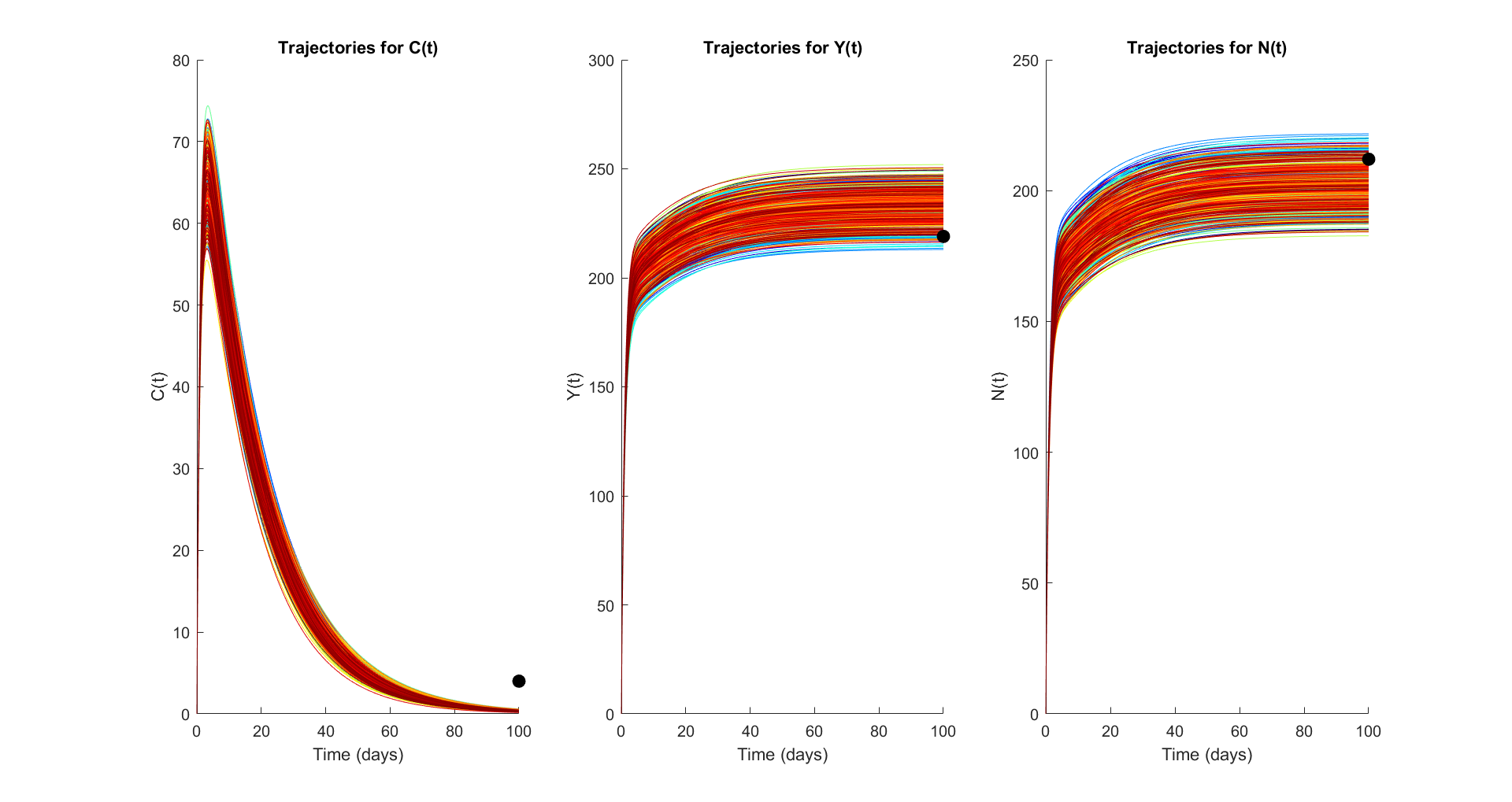} 
    \caption{This figure depicts the various trajectories of $C(t)$, $Y(t)$, and $N(t)$, produced by varying key parameters, alongside the actual vote outcomes in the House for the ACESA.}
    \label{Figure9}
\end{figure}

Further analysis reveals that the final values of $C(t)$, $Y(t)$, and $N(t)$ appear to follow a roughly normal distribution when parameters are varied. The trajectories of $C(t)$, $Y(t)$, and $N(t)$ in the House under these variations are shown in Fig.~\ref{Figure9}.

As seen in Fig.~\ref{Figure9}, the trajectories of the vote distribution often cross over the actual final votes recorded in the House. This demonstrates that a 10\% variation in key parameters can produce trajectories that reflect the real-world voting outcome, indicating that the model is robust to moderate changes in these parameters. This agreement between modeled and actual results further validates the assumptions built into the model.

\subsection{Stability and Impact of Parameter Variations}
\begin{figure}[ht!]
    \centering
    \includegraphics[width=14cm]{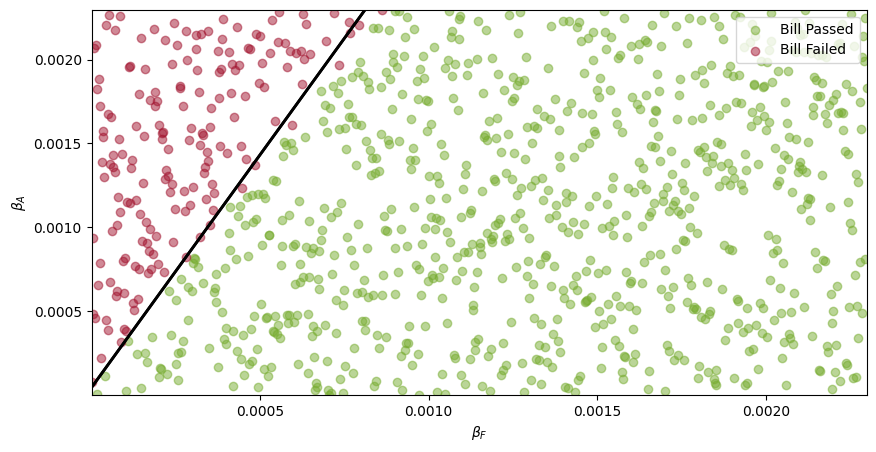}
    \caption{\textbf{ACESA results as the $\beta$ parameters vary:} This figure illustrates how variations in parameter $\beta$ affect the likelihood of the bill passing or failing. The boundaries separating pass/fail outcomes were determined using support vector machines (SVM). We use a linear SVM with a dimensionality of 2 that has been trained on 1000 samples.}
    \label{Figure10}
\end{figure}

\begin{figure}[ht!]
    \centering
    \includegraphics[width=14cm]{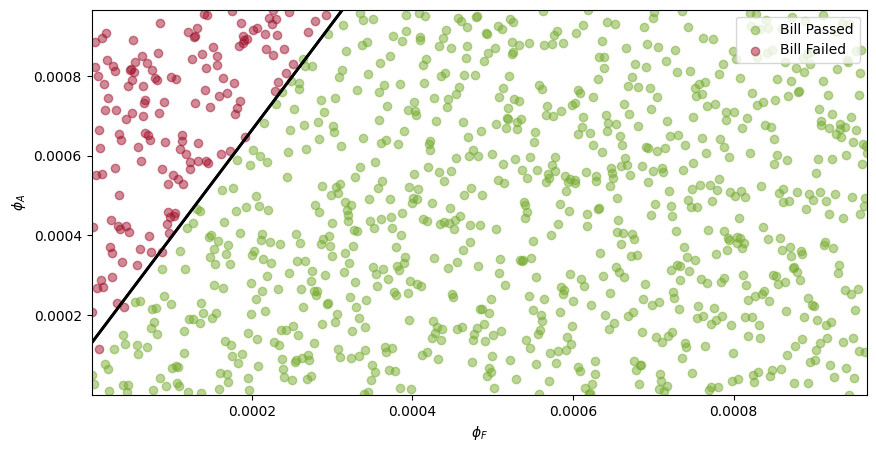}
    \caption{\textbf{ACESA results as the $\phi$ parameters vary:} This figure illustrates how variations in parameter $\phi$ affect the likelihood of the bill passing or failing. The boundaries separating pass/fail outcomes were determined using support vector machines (SVM). We use a linear SVM with a dimensionality of 2 that has been trained on 1000 samples.}
    \label{Figure11}
\end{figure}

\begin{figure}[ht!]
    \centering
    \includegraphics[width=14cm]{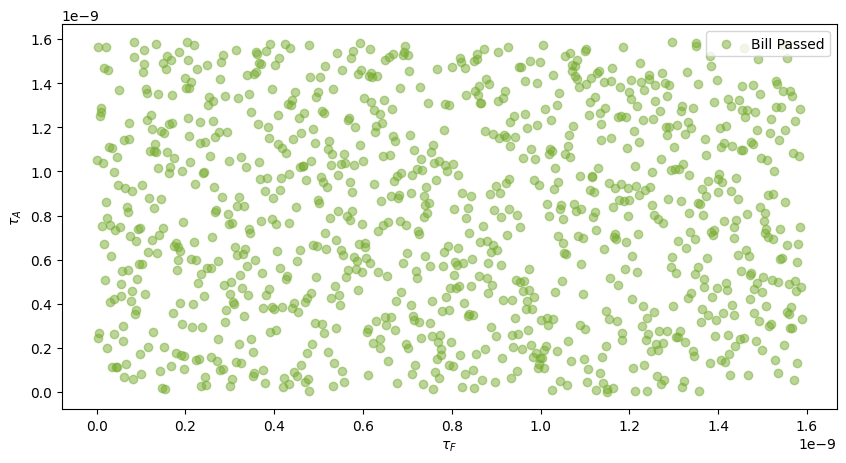}
    \caption{\textbf{ACESA results as the $\tau$ parameters vary:} This figure illustrates how variations in parameter $\tau$ affect the likelihood of the bill passing or failing. The boundaries separating pass/fail outcomes were determined using support vector machines (SVM). We use a linear SVM with a dimensionality of 2 that has been trained on 1000 samples.}
    \label{Figure12}
\end{figure}

We also explore the stability of the model by varying individual parameters controlling the flow of centrists into voting decisions. These parameters directly influence whether a legislator votes Yes or No on the bill. We examine how the bill's outcomes are influenced by variations in $\beta_F$ and $\beta_A$ in Fig.~\ref{Figure10}, by changes in $\phi_F$ and $\phi_A$ in Fig.~\ref{Figure11}, and by shifts in $\tau_F$ and $\tau_A$ in Fig.~\ref{Figure12}. Each of the red dots corresponds to a scenario where the bill fails and each of the green dots corresponds to a scenario where the bill passes.

 We applied Support Vector Machines (SVM) to identify the decision boundary between parameter sets that lead to bill passage and those that result in failure. SVMs work by mapping the input data into a higher-dimensional space and finding a hyperplane that maximizes the margin between the two classes, ensuring that the boundary is as far away as possible from the closest points in each class, known as support vectors. The SVM technique helps classify these outcomes by finding an optimal boundary that separates the two classes (pass and fail) based on the given parameter space. By applying this technique, we can determine how the parameters interact to influence the final outcome.
 
From this analysis, it becomes clear that the majority of vote outcomes suggest the bill is likely to pass under most parameter variations. Particularly, variations in $\beta$ in Fig. \ref{Figure13} and $\phi$ in Fig. \ref{Figure14} significantly impact the passage of the bill. However, $\tau$ in Fig. \ref{Figure15} does not have have a significant impact on the passage of the bill.

\subsection{Multivariate Parameter Sensitivity}

\begin{figure}[ht!]
    \centering
    \includegraphics[width=14cm]{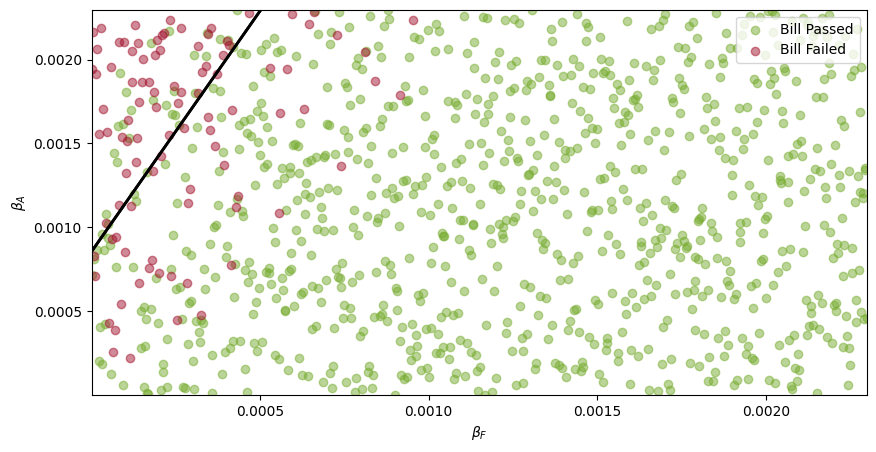}
    \caption{\textbf{ACESA results as all of the parameters vary on the $\beta_F-\beta_A$ plane:} This figure displays the results of the bill when multiple parameters, including $\beta$, are varied together. The SVM technique is used to determine the boundary between passing and failing outcomes. We use a linear SVM with a dimensionality of 6 that has been trained on 1000 samples.}
    \label{Figure13}
\end{figure}

\begin{figure}[ht!]
    \centering
    \includegraphics[width=14cm]{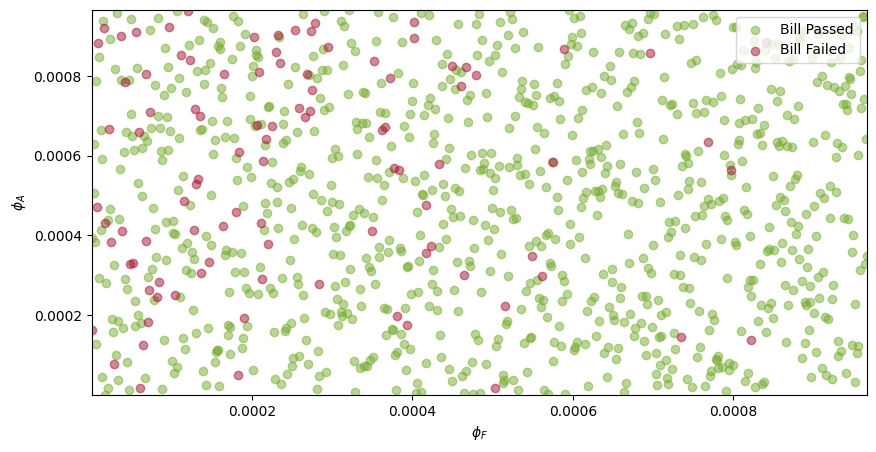}
    \caption{\textbf{ACESA results as all of the parameters vary on the $\phi_F-\phi_A$ plane:} This figure displays the results of the bill when multiple parameters, including $\phi$, are varied together. The SVM technique is used to determine the boundary between passing and failing outcomes. We use a linear SVM with a dimensionality of 6 that has been trained on 1000 samples.}
    \label{Figure14}
\end{figure}

\begin{figure}[ht!]
    \centering
    \includegraphics[width=14cm]{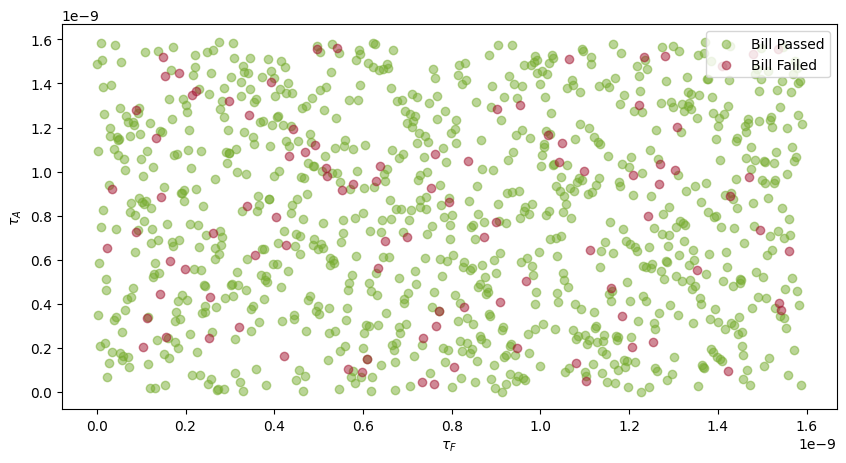}
    \caption{\textbf{ACESA results as all of the parameters vary on the $\tau_F-\tau_A$ plane:} This figure displays the results of the bill when multiple parameters, including $\tau$, are varied together. The SVM technique is used to determine the boundary between passing and failing outcomes. We use a linear SVM with a dimensionality of 6 that has been trained on 1000 samples.}
    \label{Figure15}
\end{figure}

Next, we conduct a multivariate sensitivity analysis by simultaneously varying multiple parameters. This approach allows us to observe the combined impact of interactions between different factors, offering a more comprehensive view of the system’s behavior. We analyze the changes in bill outcomes as these parameters vary, focusing on $\beta_F$ and $\beta_A$ in Fig.~\ref{Figure13}, $\phi_F$ and $\phi_A$ in Fig.~\ref{Figure14}, and $\tau_F$ and $\tau_A$ in Fig.~\ref{Figure15}.  Similar to the previous section, we use red dots to represent bills failing and green dots to represent bills passing. We also use SVMs to find the optimal boundary between bills passing and failing.

The results show that when all parameters are allowed to vary, the interactions between variables—particularly the influence of $\alpha$, $\beta$, and $\tau$ parameters—create a significant number of scenarios where the bill is predicted to pass. This is especially evident when considering the influence of centrist lobbying through $M_LF$ and $M_LA$. However, because we are varying multiple parameters, in Fig. \ref{Figure13} there are some red dots over the boundary determined by the SVM and some green dots within the red area. That happens because the beta parameter is dominated by larger values of $\phi$. We are also not able to find a boundary in the cases of $\phi$ in Fig. \ref{Figure14} and $\tau$ in Fig. \ref{Figure15} because the impact of the $\beta$ parameters creates too much overlap between the passing and failing cases.

The findings highlight the intricate interplay of numerous variables that influence the ultimate outcome of a legislative bill's progression. In complex systems such as the legislative process, even small fluctuations in one factor can disproportionately affect the final decision, leading to a scenario where marginal majorities can tip the balance one way or another. This suggests that there are no straightforward or explicit conditions that assure the passage of a bill based solely on its various rate constants, illustrating the dynamic nature of political decision-making.

In essence, the legislative environment functions much like a chaotic system, where minor changes or perturbations can manifest as significant shifts in the trajectory of the bill. The appearance of red dots among a predominately green backdrop in the data visualization is metaphorical for these unexpected changes representing moments of opposition or failure amid general support. This could be interpreted as votes or opinions that diverge from the majority consensus, thereby underscoring the fragile nature of the consensus-building process in legislative bodies.

Moreover, the underlying model is adept at encapsulating this uncertainty within a deterministic framework. While the dynamics may be governed by a set of clear rules and interactions, the outcomes remain inherently unpredictable due to the multiple layers of influence and the minor variations that act as tipping points in the decision-making process. Thus, the model accommodates the complexity of human behavior and institutional dynamics, revealing how closely held beliefs, advocacy efforts, and shifts in public opinion can sway legislative outcomes.

This nuanced understanding of bill progression is critical for stakeholders, including policymakers, analysts, and advocates, as it underscores the need for strategic approaches in navigating complex legislative environments. Identifying and targeting these key factors may provide opportunities to influence the outcome favorably. Ultimately, the proposed model not only elucidates the dynamics at play but also serves as a tool for anticipating potential results in a landscape marked by uncertainty and complexity.

\subsection{Parameter-Specific Impact}
\begin{figure}[ht!]
    \centering
    \includegraphics[width = 13cm]{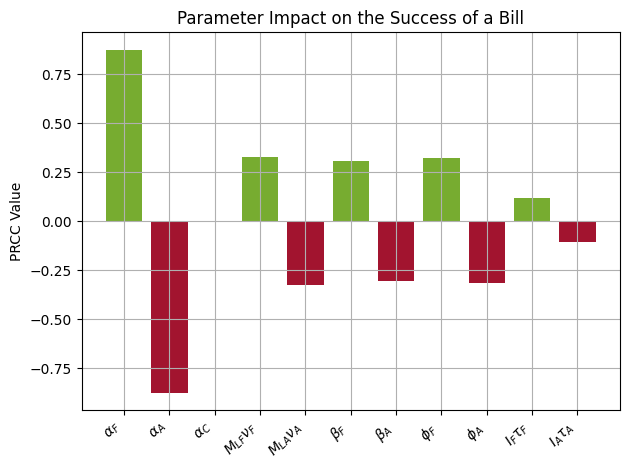}
    \caption{This figure shows the individual impact of each parameter on the outcome of whether the bill passes or fails.}
    \label{Figure16}
\end{figure}

To further isolate the contribution of each parameter to the final outcome, we perform a sensitivity analysis in Fig.~\ref{Figure16} that quantifies the individual impact of each parameter on the model’s results.

From the sensitivity analysis, it becomes evident that certain parameters, particularly those related to centrists' interactions with lobbying groups ($M_LF$, $M_LA$), exert a substantial influence on the final outcome. The influence of $\tau$, which governs the ideological pressure on decision-making, also stands out as a key factor. In contrast, parameters governing the initial size of various compartments, such as $S(t)$, exhibit minimal impact, reaffirming the idea that once legislators are involved in the decision process, the outcome is primarily shaped by the interactions modeled through other variables.

This sensitivity analysis highlights which aspects of the lobbying process and legislative dynamics are most critical to predicting the passage of the bill, offering valuable insights for both theoretical exploration and practical applications in legislative strategy. By refining our understanding of parameter influence, we can better anticipate how changes in lobbying efforts or political dynamics might affect future legislative outcomes.

\section{Results}

In this section, we validate our proposed model [1] using data on the American Clean Energy and Security Act (ACESA), commonly known as the Waxman-Markey Bill.  The model inputs are detailed in Table \ref{tab:initial_conditions} and are parameterized in Table \ref{Tab1}. Using our model, we predict both in the House and in the Senate dynamics. 

\begin{table}[h!]
    \centering
    \begin{tabular}{|c|c|c|}
        \hline
        \textbf{Variable} & \textbf{House Initial Conditions} & \textbf{Senate Initial Conditions} \\
        \( S \) & 435 & 100 \\
        \hline
        \( C \) & 0 & 0 \\
        \hline
        \( Y \) & 0 & 0 \\
        \hline
        \( N \) & 0 & 0 \\
        \hline
        \( L_F \) & 0 & 0 \\
        \hline
        \( L_A \) & 0 & 0 \\
        \hline
        \( M_F \) & 345873085 & 345873085 \\
        \hline
        \( M_A \) & 283377397 & 283377397 \\
        \hline
    \end{tabular}
    \caption{Initial Conditions for the Model in the House and Senate}
    \label{tab:initial_conditions}
\end{table}

\subsection{Dynamics in the House}

\begin{figure}[ht!]
\centering
\includegraphics[width = 10.5cm]{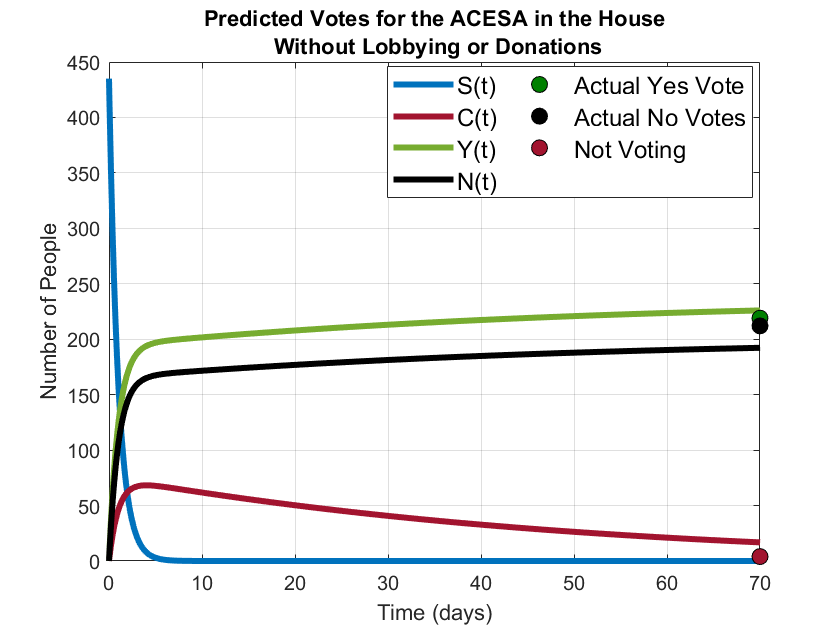}
\caption{This figure depicts the change in the opinion of legislators in the House of Representatives on the ACESA as a result of their ideology and peer pressure. As legislators are introduced to the bill, their ideology determines if they have no opinion on the bill, are in favor of it, or are against it. Then mass action peer pressure dynamics push undecided legislators to choose to vote yes or no on the bill.}
\label{Figure2}
\end{figure}

Initially, we simulate the model without considering the effects of lobbying or campaign donations to observe the voting dynamics in the U.S. House of Representatives. The results, shown in Fig. \ref{Figure2}, indicate that the model stabilizes with 230 Members voting “yes,” 196 voting “no,” and 9 remaining undecided. In comparison, the actual vote results for the bill were 219 “yes” votes, 212 “no” votes, and 4 members abstaining. As a result, the percent error for yes votes is 5.02\% and the percent error for no votes is 7.55\%.

The simulation without the influence of lobbying shows a substantial margin favoring the bill, with a significantly higher number of “yes” votes. In contrast, the actual vote outcome was much closer, suggesting that the bill’s narrow passage may have been influenced by factors not captured in the basic model without lobbying and campaign effects. The larger number of undecided votes in the model output implies that monetary influences, such as campaign donations and lobbying, may have played a role in moving undecided members toward a definitive vote.

\begin{figure}[ht!]
\centering
    \includegraphics[width = 10.5cm]{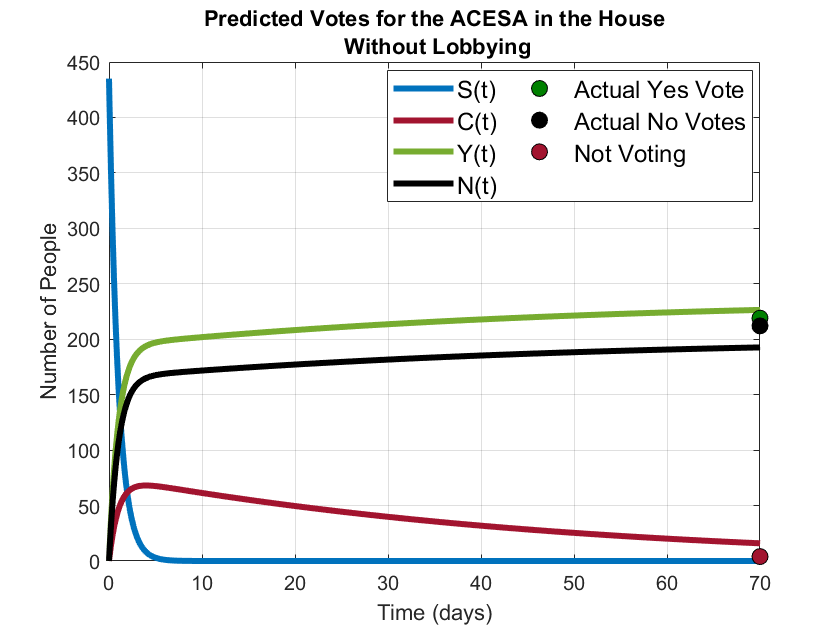}
    \caption{This figure depicts the change in the opinion of legislators in the House of Representatives on the ACESA as a result of their ideology, donations, and peer pressure. As legislators are introduced to the bill, their ideology determines if they have no opinion on the bill, are in favor of it, or are against it. Then mass action peer pressure dynamics and interactions with donations push undecided legislators to choose to vote yes or no on the bill.}
    \label{Figure3}
\end{figure}

To examine these influences, we first introduce campaign donations into the model (see Fig. \ref{Figure3}). After including campaign donations, the model results indicate 231 “yes” votes, 196 “no” votes, and 8 undecided voters, suggesting that campaign donations influenced just one undecided voter to become a “yes” voter. However, campaign donations alone did not affect members who had already made their decision, indicating that donations primarily reinforce existing positions without altering resolved decisions. Consequently, campaign donations alone appear to have limited impact on the bill’s success or failure.

\begin{figure}[ht!]
\centering
    \includegraphics[width = 10.5cm]{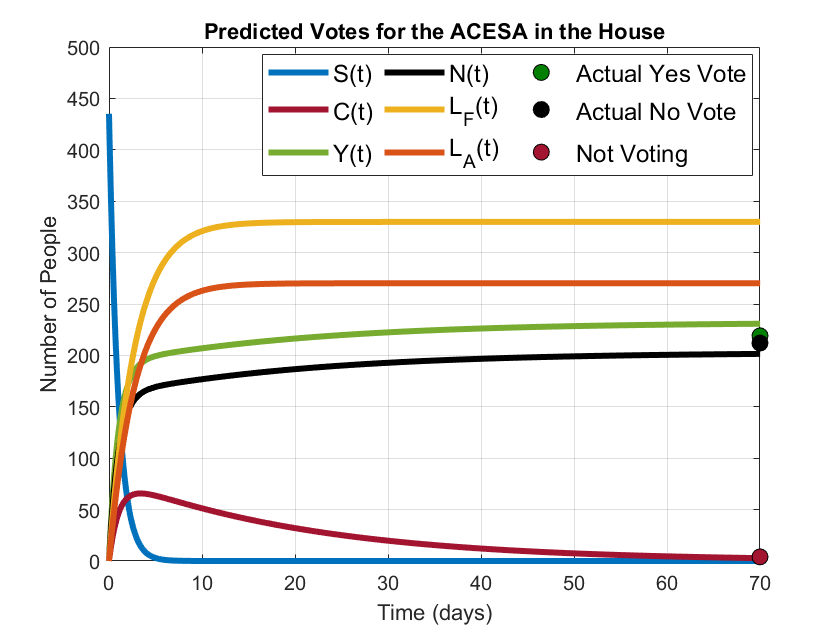}
     \caption{This figure depicts the change in the opinion of legislators in the House of Representatives on the ACESA as a result of their ideology, lobbying, financial contributions, and peer pressure. As legislators are introduced to the bill, their ideology determines if they have no opinion on the bill, are in favor of it, or are against it. Then mass action peer pressure dynamics, financial contributions, and interactions with lobbyists push undecided legislators to choose to vote yes or no on the bill.}
     \label{Figure4}
\end{figure}

Incorporating the impact of lobbying, however, brings the model’s predictions closer to the actual data. From Fig. \ref{Figure4}, we observe that with lobbying included, the model yields 231 “yes” votes, 203 “no” votes, and just 1 undecided voter, aligning more closely with the real vote count. This calculation yields a percent error of 5.48\% for the "yes" votes and 4.25\% for the "no" votes. While the percent error for the "yes" votes appears to have increased, this is primarily due to the reduction in the number of undecided voters from 9 to 1, as one of the voters who shifted from undecided to decided chose to vote "yes." This suggests that lobbying played a significant role in swaying undecided legislators to take a side, whereas campaign donations had a comparatively smaller influence on voting decisions. The impact of campaign donations may reflect legislators’ ideological leanings more than it directly influences their final votes \cite{https://doi.org/10.1111/ajps.12376}.

\subsection{Dynamics in the Senate} 
In the senate there is a rule called the filibuster that allows a senator to effectively halt any bills progression and minimum 60 votes are needed to override it. Notably, the American Clean Energy and Security Act did not achieve the same success in the Senate. The bill ultimately failed to pass there, as it could not overcome the filibuster threshold. This outcome underscores that while lobbying and campaign donations have considerable influence in the House, their effects in the Senate may be limited by procedural hurdles. We now use our model to examine this dynamic in the Senate context.

\begin{figure}[ht!]
\centering
    \includegraphics[width = 10.5cm]{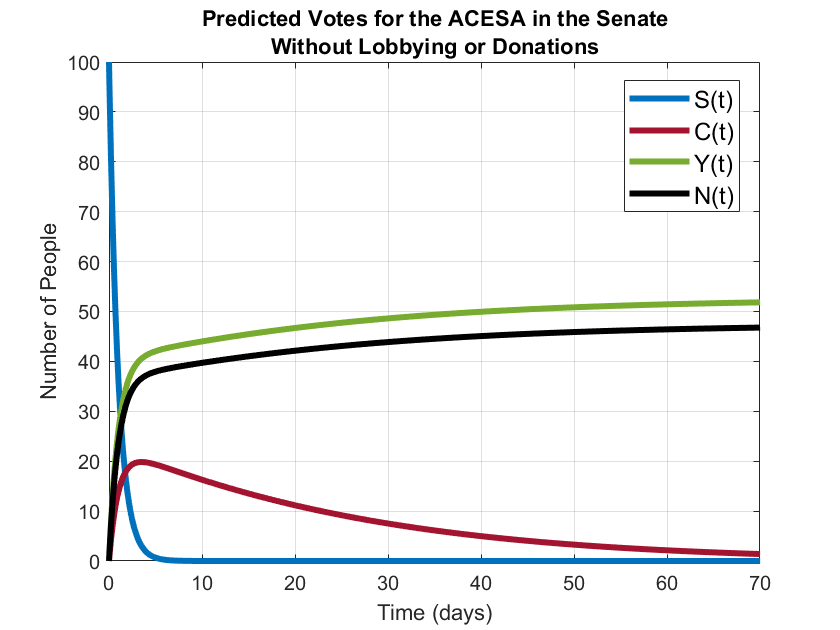}
    \caption{This figure depicts the change in the opinion of legislators in the Senate on the ACESA as a result of their ideology and peer pressure. As legislators are introduced to the bill, their ideology determines if they have no opinion on the bill, are in favor of it, or are against it. Then mass action peer pressure dynamics push undecided legislators to choose to vote yes or no on the bill.}
    \label{Figure5}
\end{figure}

\begin{figure}[ht!]
\centering
    \includegraphics[width = 10.5cm]{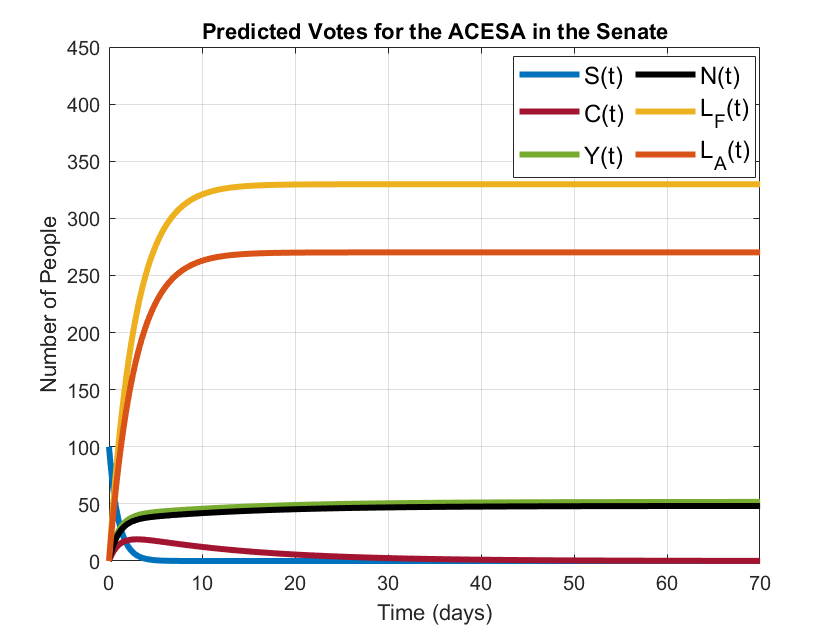}
    \caption{This figure depicts the change in the opinion of legislators in the Senate on the ACESA as a result of their ideology, lobbying, financial contributions, and peer pressure. As legislators are introduced to the bill, their ideology determines if they have no opinion on the bill, are in favor of it, or are against it. Then mass action peer pressure dynamics, financial contributions, and interactions with lobbyists push undecided legislators to choose to vote yes or no on the bill.}
    \label{Figure6}
\end{figure}

Without accounting for lobbying or campaign donations, the model predicts the Senate voting results as shown in Fig. \ref{Figure5}, with 52 “yes” votes, 47 “no” votes, and one undecided senator. This baseline simulation, focusing solely on ideological leanings, indicates a close outcome. When the impacts of lobbying and campaign donations are included in Fig. \ref{Figure6}, the model adjusts to 52 “yes” votes, 48 “no” votes, and no undecided senators, indicating a slight shift toward a more decisive outcome. Unlike in the House, lobbying appears to have minimal effect on Senate voting patterns, likely due to the strong ideological commitments of senators, which limit the influence of external pressures.

This minimal impact of lobbying in the Senate aligns with the real outcome, where the ACESA ultimately failed, falling short of the crucial 60-vote filibuster threshold. Our model effectively captures this result, demonstrating its ability to predict legislative success or failure by incorporating both ideological stances and procedural constraints. This predictive capability highlights our model's strength in evaluating the complexities of legislative voting outcomes, particularly within settings like the Senate, where restrictive procedural requirements play a decisive role.

\subsection{Impact of ideology} 
\begin{figure}[ht!]
    \centering
{\includegraphics[width=10.5cm]{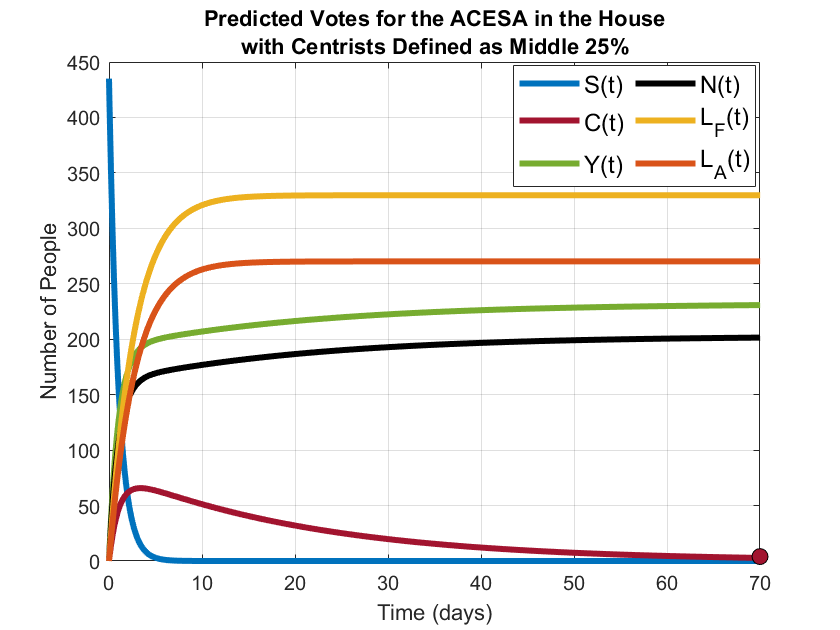} }%
\includegraphics[width=10.5cm]{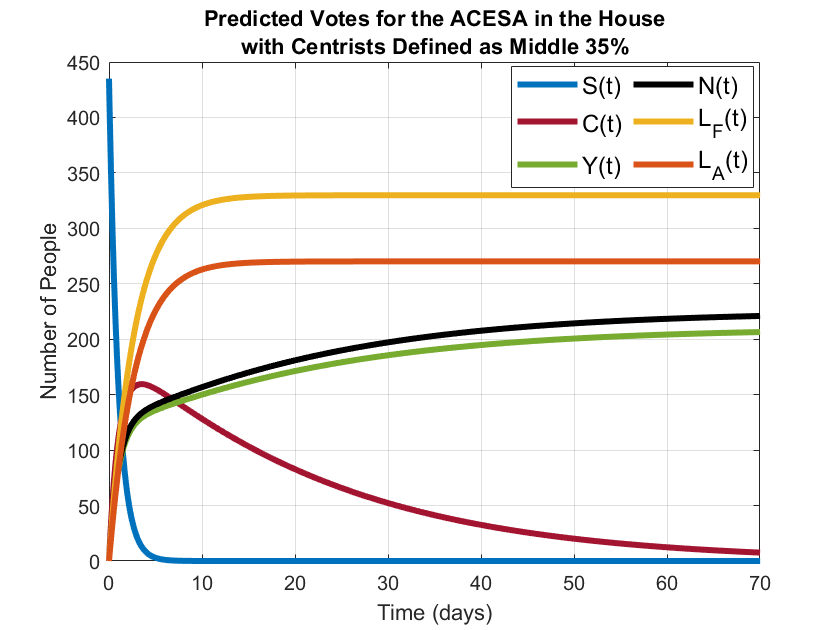} %
\caption{These figures depict the impact of different definitions of what determines an undecided voter impact the predicted result of the roll-call vote on the ACESA in the House.}
\label{Figure7}
\end{figure}

In legislative voting models, ideological distributions play a critical role, especially among centrists whose positions may shift based on external pressures like lobbying. Ideologies are measured on a spectrum, with “centrist” legislators typically identified within a central ideological range, while those with more conservative or liberal views fall outside this range. In our model, the asymptotic values of \( Y(t) \) (yes votes) and \( N(t) \) (no votes) demonstrate sensitivity to these centrist definitions, showing the model’s adaptability to changes in ideological influence. Initially, centrists were defined as those within $\pm 25\%$ of true centrist (0) on the ideological scale. Legislators with ideologies between -0.25 and 0.25 on a scale from -1 (most liberal) to 1 (most conservative) were categorized as centrist and assigned to the C compartment. Expanding the centrist range to include ideologies from -0.35 to 0.35, as shown in Figure \ref{Figure7}, shifts the outcome, resulting in “no” votes exceeding “yes” votes.

This variability based on centrist definitions highlights how even slight changes in ideological distribution can impact the model’s predicted voting outcomes. It emphasizes the role that ideological proximity and influence play in shaping legislative outcomes, as senators near the center may be more susceptible to both ideological and external pressures. Therefore, these results not only capture the nuanced dynamics of voting behavior but also demonstrate how shifts in influence among centrists can affect legislative success or failure, reflecting real-world complexities seen in the ACESA’s progression. 


\section{Discussion}
In this study, we’ve developed a novel compartmental model to predict legislative outcomes, particularly the passage of bills, marking, to our knowledge, the first use of this approach in political science. Validated against the voting results for the American Clean Energy and Security Act (ACESA), the model accurately simulates voting dynamics, underscoring how factors like lobbying, peer pressure, and ideological composition drive legislative decisions. By simulating various scenarios, this model provides legislators with the strategic advantage to refine bill proposals and helps lobbyists identify where to focus their influence on undecided legislators.

This research introduces a pioneering approach in using compartmental models to predict legislative outcomes, with promising results validated by ACESA data. By enabling continuous analysis of legislative dynamics, our model captures how lobbying, peer influence, and ideological composition collectively shape voting behavior. This offers policymakers and advocates a tool to navigate the complexities of legislative strategy more effectively.

Applying real-time data and stochastic elements will only enhance the model’s accuracy and relevance, allowing it to mirror the unpredictable yet patterned nature of legislative decision-making. This capability is essential for those invested in understanding the political landscape in real-time, where strategic adaptations can determine legislative success.

Ultimately, this study lays foundational work in political science modeling, enabling the use of continuous, dynamic models that enrich our understanding of legislative processes. This approach not only brings rigor to legislative strategy but also opens doors for future research that will expand how we analyze, predict, and ultimately shape legislative outcomes.

Our stability analysis shows that convergence in voting decisions over time is driven by the combined effects of peer pressure, lobbying, donations, and ideological distribution, suggesting that legislative bodies may tend toward stable outcomes under certain pressures. This tendency to stabilize aligns with real-world observations of legislative behavior, where peer influence and ideological cohesion can lead to predictable voting patterns.


To enhance future research on legislative behavior, it is essential to broaden the scope of bills analyzed using the proposed model. By applying this model to a diverse array of legislation, researchers can systematically assess its robustness and adaptability across different contexts and subject matters. This expansion not only helps to validate the initial findings but also uncovers nuanced relationships and patterns that may only emerge in specific legislative environments.

Incorporating advanced techniques for managing complex, interacting variables is pivotal in this endeavor. Legislative behavior is inherently multifaceted, influenced by numerous factors such as political ideology, constituent interests, party dynamics, and socio-economic conditions. By adopting suitable analytical methods—such as multilevel modeling, machine learning algorithms, or network analysis—researchers can better capture the idynamics of these interactions. This could involve exploring how different factors simultaneously impact legislative decision-making or identifying potential feedback loops within the political system.

Moreover, employing interdisciplinary approaches will further enable the framework for understanding legislative behavior. Drawing insights from political science, sociology, economics, and data science can provide a more holistic view of the dynamics at play. For example, insights from behavioral economics could inform how cognitive biases influence legislators' choices, while sociological perspectives might offer valuable context regarding the influence of social networks and group dynamics.

\section{Acknowledgement} The corresponding author is
 grateful to AMS-Simons Research Enhancement Grant for providing the resources and facilities necessary for my research. Their contributions were vital in allowing me to conduct my work effectively.

\vspace{10em}
\newpage

\bibliographystyle{plain}
\bibliography{references.bib}

\begin{thebibliography}{10}

\bibitem{aragon2024mathematical}
Francisco~J Arag{\'o}n-Artacho and Miguel~A Goberna.
\newblock Mathematical tools for political decision-making.
\newblock In {\em Mathematics in Politics and Governance}, pages 1--42. Springer, 2024.

\bibitem{aragon2024mathematics}
Francisco~J Arag{\'o}n-Artacho and Miguel~A Goberna.
\newblock {\em Mathematics in Politics and Governance}.
\newblock Springer, 2024.

\bibitem{Bari2021}
Anasse Bari, William Brower, and Christopher Davidson.
\newblock Using artificial intelligence to predict legislative votes in the united states congress.
\newblock In {\em 2021 IEEE 6th International Conference on Big Data Analytics (ICBDA)}, pages 56--60, 2021.

\bibitem{Money1}
Frank~R. Baumgartner, Jeffrey~M. Berry, Marie Hojnacki, David~C. Kimball, and Beth~L. Leech.
\newblock Money, priorities, and stalemate: How lobbying affects public policy.
\newblock {\em Election Law Journal: Rules, Politics, and Policy}, 13(1):194--209, 2014.

\bibitem{https://doi.org/10.1111/ajps.12376}
Adam Bonica.
\newblock Inferring roll-call scores from campaign contributions using supervised machine learning.
\newblock {\em American Journal of Political Science}, 62(4):830--848, 2018.

\bibitem{Budhwar2018}
Aditya Budhwar, Toshihiro Kuboi, Alex Dekhtyar, and Foaad Khosmood.
\newblock Predicting the vote using legislative speech.
\newblock In {\em Proceedings of the 19th Annual International Conference on Digital Government Research: Governance in the Data Age}, dg.o '18, New York, NY, USA, 2018. Association for Computing Machinery.

\bibitem{doi:10.1086/705816}
Christian Fong.
\newblock Expertise, networks, and interpersonal influence in congress.
\newblock {\em The Journal of Politics}, 82(1):269--284, 2020.

\bibitem{c2es2024}
Center for Climate and Energy Solutions.
\newblock Congress climate history - center for climate and energy solutions, 2024.
\newblock Accessed January 17, 2024.

\bibitem{Goldblatt2012}
David Goldblatt and Tyler O’Neil.
\newblock How a bill becomes a law - predicting votes from legislation text.
\newblock Technical report, Stanford University, 2012.

\bibitem{HenighanKravitz}
Tom Henighan and Scott Kravitz.
\newblock Predicting bill votes in the house of representatives.

\bibitem{https://doi.org/10.1111/fima.12032}
Matthew~D. Hill, G.~Wayne Kelly, G.~Brandon Lockhart, and Robert~A. Van~Ness.
\newblock Determinants and effects of corporate lobbying.
\newblock {\em Financial Management}, 42(4):931--957, 2013.

\bibitem{adbc7077-a656-3db1-a131-059742c1c49c}
Marie Hojnacki and David~C. Kimball.
\newblock Organized interests and the decision of whom to lobby in congress.
\newblock {\em The American Political Science Review}, 92(4):775--790, 1998.

\bibitem{Karimi2020}
Hamid Karimi, Tyler Derr, Aaron Brookhouse, and Jiliang Tang.
\newblock Multi-factor congressional vote prediction.
\newblock In {\em Proceedings of the 2019 IEEE/ACM International Conference on Advances in Social Networks Analysis and Mining}, ASONAM '19, page 266–273, New York, NY, USA, 2020. Association for Computing Machinery.

\bibitem{KimKunisky2021}
In~Song Kim and Dmitriy Kunisky.
\newblock Mapping political communities: A statistical analysis of lobbying networks in legislative politics.
\newblock {\em Political Analysis}, 29(3):317–336, 2021.

\bibitem{Lan2013}
JY~Lan and S~Shah.
\newblock Utilizing network analysis to model congressional voting behavior.
\newblock {\em Computer Science}, 2013.

\bibitem{voteview}
Jeffrey~B. Lewis, Keith Poole, Howard Rosenthal, Adam Boche, Aaron Rudkin, and Luke Sonnet.
\newblock Voteview: Congressional roll-call votes database, 2023.

\bibitem{meng2019}
Kyle~C. Meng and Ashwin Rode.
\newblock The social cost of lobbying over climate policy.
\newblock {\em Nature Climate Change}, 9:472--476, 2019.

\bibitem{nehama2015complexity}
Ilan Nehama.
\newblock Complexity of optimal lobbying in threshold aggregation.
\newblock In {\em Algorithmic Decision Theory: 4th International Conference, ADT 2015, Lexington, KY, USA, September 27--30, 2015, Proceedings 4}, pages 379--395. Springer, 2015.

\bibitem{Poole1985}
Keith~T. Poole and Howard Rosenthal.
\newblock A spatial model for legislative roll call analysis.
\newblock {\em American Journal of Political Science}, 29(2):357--384, 1985.

\bibitem{raissi2019parameter}
Maziar Raissi, Niloofar Ramezani, and Padmanabhan Seshaiyer.
\newblock On parameter estimation approaches for predicting disease transmission through optimization, deep learning and statistical inference methods.
\newblock {\em Letters in biomathematics}, 6(2):1--26, 2019.

\bibitem{Smith2012}
Samuel Smith, Jae~Yeon Baek, Zhaoyi Kang, Dawn Song, Laurent~El Ghaoui, and Mario Frank.
\newblock Predicting congressional votes based on campaign finance data.
\newblock In {\em 2012 11th International Conference on Machine Learning and Applications}, volume~1, pages 640--645, 2012.

\bibitem{stoddard2021}
Isak Stoddard, Kevin Anderson, Stuart Capstick, Joanna Depledge, Keri Facer, and Clair Gough.
\newblock Three decades of climate mitigation: Why haven’t we bent the global emissions curve?
\newblock {\em Annual Review of Environment and Resources}, 46:653--689, 2021.

\bibitem{ward2004pressure}
Hugh Ward.
\newblock Pressure politics: A game-theoretical investigation of lobbying and the measurement of power.
\newblock {\em Journal of theoretical politics}, 16(1):31--52, 2004.

\bibitem{Willis2024}
Derek Willis, Annie McCartney, and Jamie~B. Merrill.
\newblock Represent.
\newblock \url{https://projects.propublica.org/represent/}, 2024.
\newblock Accessed: 20 June 2024.

\end{thebibliography}

\newpage
\section{Appendix}
\label{Appendix}

\subsection{Stability of the Lobbying and Donations-Free Model}
Consider the corresponding model system:

\begin{subequations}\label{reduced_model}
\begin{eqnarray}
\frac{dS}{dt} &=& -(\alpha_F + \alpha_A + \alpha_C)S, \\
\frac{dC}{dt} &=& \alpha_C S - \beta_F CY - \beta_A CN, \\
\frac{dY}{dt} &=& \alpha_F S + \beta_F CY, \\
\frac{dN}{dt} &=& \alpha_A S + \beta_A CN.
\end{eqnarray}
\end{subequations}

To determine the stability of the equilibrium point \((S,C,Y,N) = (0, C_0, 0, 0)\), we use the transformations \(S(t) = s(t)\), \(C(t) = C_0 + c(t)\), \(Y(t) = y(t)\), and \(N(t) = n(t)\). Substituting these transformations into \eqref{reduced_model} and linearizing the system by neglecting higher-order terms of the small quantities \(s(t)\), \(c(t)\), \(y(t)\), and \(n(t)\), we obtain the following system of linear equations:

\begin{subequations}\label{linear_equations}
\begin{eqnarray}
\frac{ds}{dt} &=& -(\alpha_F + \alpha_A + \alpha_C)s, \\
\frac{dc}{dt} &=& \alpha_C s - \beta_F C_0 y - \beta_A C_0 n, \\
\frac{dy}{dt} &=& \alpha_F s + \beta_F C_0 y, \\
\frac{dn}{dt} &=& \alpha_A s + \beta_A C_0 n.
\end{eqnarray}
\end{subequations}

This system can be written in matrix form as:

\[
\frac{d}{dt} \begin{bmatrix}
    s \\
    c \\
    y \\
    n
\end{bmatrix} = \begin{bmatrix}
    -(\alpha_F + \alpha_A + \alpha_C) & 0 & 0 & 0 \\
    \alpha_C & 0 & -\beta_F C_0 & -\beta_A C_0 \\
    \alpha_F & 0 & \beta_F C_0 & 0 \\
    \alpha_A & 0 & 0 & \beta_A C_0
\end{bmatrix} \begin{bmatrix}
    s \\
    c \\
    y \\
    n
\end{bmatrix}.
\]
To simplify the calculations, we use a transformation:

\[
\begin{bmatrix}
    s \\
    c \\
    y \\
    n
\end{bmatrix} = \begin{bmatrix}
    p_{11} & p_{12} & p_{13} & p_{14} \\
    p_{21} & p_{22} & p_{23} & p_{24} \\
    p_{31} & p_{32} & p_{33} & p_{34} \\
    p_{41} & p_{42} & p_{43} & p_{44}
\end{bmatrix} \begin{bmatrix}
    s_1 \\
    c_1 \\
    y_1 \\
    n_1
\end{bmatrix}
\]

 where $p_{ij}$ represents the corresponding eigenvalues of the matrix. We obtain the following system:

\[
\frac{d}{dt} \begin{bmatrix}
    s_1 \\
    c_1 \\
    y_1 \\
    n_1
\end{bmatrix} = \begin{bmatrix}
    -(\alpha_F + \alpha_A + \alpha_C) & 0 & 0 & 0 \\
    0 & 0 & 0 & 0 \\
    0 & 0 & \beta_F C_0 & 0 \\
    0 & 0 & 0 & \beta_A C_0
\end{bmatrix} \begin{bmatrix}
    s_1 \\
    c_1 \\
    y_1 \\
    n_1
\end{bmatrix}.
\]

The solution is:

\[
\begin{bmatrix}
    s_1(t) \\
    c_1(t) \\
    y_1(t) \\
    n_1(t)
\end{bmatrix} = \begin{bmatrix}
    s_{10} e^{-(\alpha_F + \alpha_A + \alpha_C)t} \\
    c_{10} \\
    y_{10} e^{\beta_F C_0 t} \\
    n_{10} e^{\beta_A C_0 t}
\end{bmatrix}.
\]

From these solutions, we conclude that as \(t \to \infty\), \(s(t) \to 0\), \(y(t) \to +\infty\), etc. Thus, the equilibrium point \((0, C_0, 0, 0)\) is unstable.
This implies that when the centrist group is dominant ($C_0$) and the "yes" ($Y$) and "no" ($N$) voting groups are negligible, the situation is inherently unstable. Even minor influences, such as campaigns or debates, can disrupt this balance, prompting individuals to take a stance, either in favor of ($Y$) or against ($N$) the bill. As a result, the population is unlikely to remain neutral.

A similar calculation can be used to determine the stability of the equilibrium point \((S, C, Y, N) = (0, 0, Y_0, N_0)\). Using the transformations \(S(t) = s(t)\), \(C(t) = c(t)\), \(Y(t) = Y_0 + y(t)\), and \(N(t) = N_0 + n(t)\), we linearize \eqref{reduced_model}:

\[
\frac{d}{dt} \begin{bmatrix}
    s \\
    c \\
    y \\
    n
\end{bmatrix} = \begin{bmatrix}
    -(\alpha_F + \alpha_A + \alpha_C) & 0 & 0 & 0 \\
    \alpha_C & -\beta_F Y_0 - \beta_A N_0 & 0 & 0 \\
    \alpha_F & \beta_F Y_0 & 0 & 0 \\
    \alpha_A & \beta_A N_0 & 0 & 0
\end{bmatrix} \begin{bmatrix}
    s \\
    c \\
    y \\
    n
\end{bmatrix}.
\]

Transforming the system, we obtain:

\[
\begin{bmatrix}
    s \\
    c \\
    y \\
    n
\end{bmatrix} = \begin{bmatrix}
    p_{11} & p_{12} & p_{13} & p_{14} \\
    p_{21} & p_{22} & p_{23} & p_{24} \\
    p_{31} & p_{32} & p_{33} & p_{34} \\
    p_{41} & p_{42} & p_{43} & p_{44}
\end{bmatrix} \begin{bmatrix}
    s_1 \\
    c_1 \\
    y_1 \\
    n_1
\end{bmatrix},
\]

which simplifies to:

\[
\frac{d}{dt} \begin{bmatrix}
    s_1 \\
    c_1 \\
    y_1 \\
    n_1
\end{bmatrix} = \begin{bmatrix}
    -(\alpha_F + \alpha_A + \alpha_C) & 0 & 0 & 0 \\
    0 & -\beta_F Y_0 - \beta_A N_0 & 0 & 0 \\
    0 & 0 & 0 & 0 \\
    0 & 0 & 0 & 0
\end{bmatrix} \begin{bmatrix}
    s_1 \\
    c_1 \\
    y_1 \\
    n_1
\end{bmatrix}.
\]

The solution is:

\[
\begin{bmatrix}
    s_1(t) \\
    c_1(t) \\
    y_1(t) \\
    n_1(t)
\end{bmatrix} = \begin{bmatrix}
    s_{10} e^{-(\alpha_F + \alpha_A + \alpha_C)t} \\
    c_{10} e^{-(\beta_F Y_0 + \beta_A N_0)t} \\
    y_{10} \\
    n_{10}
\end{bmatrix}.
\]

Transforming back to the original coordinates, we conclude that as \(t \to \infty\), \(s(t) \to 0\), \(c(t) \to 0\), etc. Thus, \((0, 0, Y_0, N_0)\) is stable. From this result, we conclude that as \(t \to \infty\), the \(Y(t)\) and \(N(t)\) functions will converge to positive constants, while the \(S(t)\) and \(C(t)\) functions will approach zero. 

\subsubsection{Solution to the \texorpdfstring{$L_F(t)$ and $L_A(t)$}{LF(t) and LA(t)} Functions}

Given the differential equations:

\begin{align*}
\frac{dL_F}{dt} &= \nu_F M_{LF}(L - L_F(t) - L_A(t)), \\
\frac{dL_A}{dt} &= \nu_A M_{LA}(L - L_F(t) - L_A(t)),
\end{align*}

we rewrite them in a more suitable form below for analysis.




\begin{align*}
\frac{dL_F}{dt} &= \nu_F M_{LF} X(t), \\
\frac{dL_A}{dt} &= \nu_A M_{LA} X(t).
\end{align*}

Notice that:

\[
X(t) = L - L_F(t) - L_A(t).
\]

Differentiating \(X(t)\):

\[
\frac{dX}{dt} = -\frac{dL_F}{dt} - \frac{dL_A}{dt}.
\]

Substitute the original equations into this:

\[
\frac{dX}{dt} = -(\nu_F M_{LF} + \nu_A M_{LA}) X(t).
\]

Simplify:

\[
\frac{dX}{dt} = -(\nu_F M_{LF} + \nu_A M_{LA}) X(t).
\]

This is a first-order linear differential equation:

\[
\frac{dX}{dt} = -(\nu_F M_{LF} + \nu_A M_{LA}) X(t).
\]

The solution to this equation is:

\[
X(t) = X(0) e^{-(\nu_F M_{LF} + \nu_A M_{LA}) t}.
\]

Since \(X(0) = L - L_F(0) - L_A(0)\), we have:

\[
X(t) = (L - L_F(0) - L_A(0)) e^{-(\nu_F M_{LF} + \nu_A M_{LA}) t}.
\]

Using the expressions for \(\frac{dL_F}{dt}\) and \(\frac{dL_A}{dt}\):

\begin{align*}
\frac{dL_F}{dt} &= \nu_F M_{LF} X(t), \\
\frac{dL_A}{dt} &= \nu_A M_{LA} X(t).
\end{align*}

Since we have \(X(t)\), we can integrate to find \(L_F(t)\) and \(L_A(t)\).

Integrate both sides:

\[
L_F(t) = L_F(0) + \nu_F M_{LF} \int_0^t X(\tau) \, d\tau,
\]

\[
L_A(t) = L_A(0) + \nu_A M_{LA} \int_0^t X(\tau) \, d\tau.
\]

Substitute \(X(\tau)\):

\begin{align*}
L_F(t) &= L_F(0) + \nu_F M_{LF} (L - L_F(0) - L_A(0)) \int_0^t e^{-(\nu_F M_{LF} + \nu_A M_{LA}) \tau} \, d\tau, \\
L_A(t) &= L_A(0) + \nu_A M_{LA} (L - L_F(0) - L_A(0)) \int_0^t e^{-(\nu_F M_{LF} + \nu_A M_{LA}) \tau} \, d\tau.
\end{align*}

Integrate:

\[
\int_0^t e^{-(\nu_F M_{LF} + \nu_A M_{LA}) \tau} \, d\tau = \frac{1 - e^{-(\nu_F M_{LF} + \nu_A M_{LA}) t}}{\nu_F M_{LF} + \nu_A M_{LA}}.
\]

So, we get:

\begin{align*}
L_F(t) &= L_F(0) + \nu_F M_{LF} (L - L_F(0) - L_A(0)) \frac{1 - e^{-(\nu_F M_{LF} + \nu_A M_{LA}) t}}{\nu_F M_{LF} + \nu_A M_{LA}}, \\
L_A(t) &= L_A(0) + \nu_A M_{LA} (L - L_F(0) - L_A(0)) \frac{1 - e^{-(\nu_F M_{LF} + \nu_A M_{LA}) t}}{\nu_F M_{LF} + \nu_A M_{LA}}.
\end{align*}

Simplify:

\begin{align*}
L_F(t) &= L_F(0) + \frac{\nu_F M_{LF}}{\nu_F M_{LF} + \nu_A M_{LA}} (L - L_F(0) - L_A(0)) (1 - e^{-(\nu_F M_{LF} + \nu_A M_{LA}) t}), \\
L_A(t) &= L_A(0) + \frac{\nu_A M_{LA}}{\nu_F M_{LF} + \nu_A M_{LA}} (L - L_F(0) - L_A(0)) (1 - e^{-(\nu_F M_{LF} + \nu_A M_{LA}) t}).
\end{align*}
\subsubsection{Full Stability Analysis}

To determine the stability of the equilibrium point \((S, C, Y, N) = (0, 0, Y_0, N_0)\), we use the transformations \(S(t) = s(t)\), \(C(t) = c(t)\), \(Y(t) = Y_0 + y(t)\), and \(N(t) = N_0 + n(t)\) and linearize the system \eqref{full_system} by neglecting higher-order terms of the small quantities \(s(t)\), \(c(t)\), \(y(t)\), \(n(t)\). The functions \(L_F(t)\) and \(L_A(t)\) are given by the previously derived solutions to their differential equations:
\small
\[
\frac{d}{dt} \begin{bmatrix}
    s \\
    c \\
    y \\
    n
\end{bmatrix} = \begin{bmatrix}
    -(\alpha_F + \alpha_A + \alpha_C) & 0 & 0 & 0 \\
    \alpha_C & -(\beta_F Y_0 + \beta_A N_0 + \phi_F L_F(t) + & 0 & 0 \\
    &\phi_A L_A(t) + \tau_F I_F(t) + \tau_A I_A(t)) &  &  \\
    \alpha_F & 0 & \beta_F Y_0 + & 0\\
   & & \phi_F L_F(t) + \tau_F I_F(t) &  \\
    \alpha_A & 0 & 0 & \beta_A N_0 + \\
    & & & \phi_A L_A(t) + \tau_A I_A(t)
\end{bmatrix} \begin{bmatrix}
    s \\
    c \\
    y \\
    n
\end{bmatrix}.
\]

Transforming the system, we obtain:

\[
\begin{bmatrix}
    s \\
    c \\
    y \\
    n
\end{bmatrix} = \begin{bmatrix}
    p_{11} & p_{12} & p_{13} & p_{14} \\
    p_{21} & p_{22} & p_{23} & p_{24} \\
    p_{31} & p_{32} & p_{33} & p_{34} \\
    p_{41} & p_{42} & p_{43} & p_{44}
\end{bmatrix} \begin{bmatrix}
    s_1 \\
    c_1 \\
    y_1 \\
    n_1
\end{bmatrix},
\]

which simplifies to:

\[
\frac{d}{dt} \begin{bmatrix}
    s_1 \\
    c_1 \\
    y_1 \\
    n_1
\end{bmatrix} = \begin{bmatrix}
    -(\alpha_F + \alpha_A + \alpha_C) & 0 & 0 & 0 \\
    0 & -(\beta_F Y_0 + \beta_A N_0 + \phi_F L_F(t) + \phi_A L_A(t) + \tau_F I_F(t) + \tau_A I_A(t)) & 0 & 0 \\
    0 & 0 & 0 & 0 \\
    0 & 0 & 0 & 0
\end{bmatrix} \begin{bmatrix}
    s_1 \\
    c_1 \\
    y_1 \\
    n_1
\end{bmatrix}.
\]

The solution to this system is:

\[
\begin{bmatrix}
    s_1(t) \\
    c_1(t) \\
    y_1(t) \\
    n_1(t)
\end{bmatrix} = \begin{bmatrix}
    s_{10} e^{-(\alpha_F + \alpha_A + \alpha_C)t} \\
    c_{10} e^{\int \left( -(\beta_F Y_0 + \beta_A N_0 + \phi_F L_F(t) + \phi_A L_A(t) + \tau_F I_F(t) + \tau_A I_A(t)) \right) dt} \\
    y_{10} \\
    n_{10}
\end{bmatrix}.
\]

\begin{longtable}{|c|p{3.5cm}|c|c|c|}
\caption{Parameter values} \label{Tab1} \\
\hline
\textbf{Parameters} & \textbf{Description} & \textbf{Units} & \textbf{Value} & \textbf{Source} \\
\hline
\endfirsthead

\multicolumn{5}{c}%
{{\tablename\ \thetable{} -- continued from previous page}} \\
\hline
\textbf{Parameters} & \textbf{Description} & \textbf{Units} & \textbf{Value} & \textbf{Source} \\
\hline
\endhead

\multicolumn{5}{|r|}{{Continued on next page}} \\ \hline
\endfoot

\hline
\endlastfoot

$\alpha_{FH}$ & Rate at which the legislators turn in favour of the bill in the House & $time^{-1}$ & 0.449 & \cite{voteview} \\
$\alpha_{AH}$ & Rate at which the legislators turn against the bill in the House & $time^{-1}$ & 0.382 & \cite{voteview} \\
$\alpha_{CH}$ & Rate at which the legislators becomes centrists in the House & $time^{-1}$ & 0.168 & \cite{voteview} \\
$\alpha_{FS}$ & Rate at which the legislators turn in favour of the bill in the Senate & $time^{-1}$ & 0.41 & \cite{voteview} \\
$\alpha_{AS}$ & Rate at which the legislators turn against the bill in the Senate & $time^{-1}$ & 0.37 & \cite{voteview} \\
$\alpha_{CS}$ & Rate at which the legislators becomes centrists in the Senate & $time^{-1}$ & 0.22 & \cite{voteview} \\
$\nu$ & Rate at which the lobbyists turn in favour of the bill & $time^{-1} money^{-1}$ & 0.0000000005258143724 & \cite{Willis2024} \\
$\psi$ & Percent of money invested in campaigning and directly paying the neutral individuals & $time^{-1}$ & 0.47847 & \cite{Money1} \\
$I_F(t)$ & Rate of outflow of money in campaign contributions in favour of the bill & $money\; time^{-1}$ & 345873.09 & \cite{meng2019} \\
$I_A(t)$ & Rate of outflow of money in campaign contributions against the bill & $money\; time^{-1}$ & 283377.340 & \cite{meng2019} \\
$\beta_{FH}$ & Rate at which centrists become in favour of the bill due to interaction in the House & $time^{-1}people^{-1}$ & 0.00005425 & \cite{doi:10.1086/705816} \\
$\beta_{AH}$ & Rate at which centrists become against the bill due to interaction in the House & $time^{-1}people^{-1}$ & 0.00005425 & \cite{doi:10.1086/705816} \\
$\beta_{FS}$ & Rate at which centrists become in favour of the bill due to interaction in the Senate & $time^{-1}people^{-1}$ & 0.000437 & \cite{doi:10.1086/705816} \\
$\beta_{AS}$ & Rate at which centrists become against the bill due to interaction in the Senate & $time^{-1}people^{-1}$ & 0.000437 & \cite{doi:10.1086/705816} \\
$\phi_{FS}$ & Rate at which centrists change their decision due to interaction with lobbyists & $time^{-1}people^{-1}$ & 0.00005117586796 & \cite{Money1} \cite{adbc7077-a656-3db1-a131-059742c1c49c} \cite{meng2019} \\
$\phi_{AS}$ & Rate at which centrists change their decision due to interaction with lobbyists & $time^{-1}people^{-1}$ & 0.00007585714286 & \cite{Money1} \cite{adbc7077-a656-3db1-a131-059742c1c49c} \cite{meng2019} \\
$\phi_{FH}$ & Rate at which centrists change their decision due to interaction with lobbyists & $time^{-1}people^{-1}$ & 0.00003461169813 & \cite{Money1} \cite{adbc7077-a656-3db1-a131-059742c1c49c} \cite{meng2019} \\
$\phi_{AH}$ & Rate at which centrists change their decision due to interaction with lobbyists & $time^{-1}people^{-1}$ & 0.00005130434783 & \cite{Money1} \cite{adbc7077-a656-3db1-a131-059742c1c49c} \\
$\tau$ & rate at which centrists become against the bill due to campaigning & $money^{-1}$ & 000000001112433819 & \cite{Money1}  \\
$L$& Max number of lobbyists that can be involved in bill & $people$  & 600 &  \cite{Willis2024} \cite{{https://doi.org/10.1111/fima.12032}}\\
\hline
\end{longtable}

\end{document}